\documentclass[a4paper,11pt]{article}
\pdfoutput=1
\usepackage{subcaption}
\usepackage[export]{adjustbox}
\usepackage{amsmath}
\usepackage[utf8]{inputenc}
\usepackage[T1]{fontenc}
\usepackage{algorithm,algpseudocode}
\usepackage{jheppub}
\usepackage{subfiles}
\usepackage{xcolor}
\usepackage{enumerate}
\usepackage[section]{placeins}
\graphicspath{ {figures/} }

\newcommand{\qgraf}{\texttt{Qgraf}}
\newcommand{\reduze}{\texttt{Reduze\,2}}
\newcommand{\form}{\texttt{FORM}}
\newcommand{\pysecdec}{\texttt{pySecDec}}
\newcommand{\singular}{\texttt{Singular}}
\newcommand{\ud}{\mathrm{d}}
\makeatletter
\renewcommand\@fpheader{}
\renewcommand\@journal{}
\makeatother
\title{\boldmath Two-loop helicity amplitudes for $gg\to ZZ$ with full top-quark mass effects}
\preprint{
\hphantom{.}\hfill CERN-TH-2020-201\\
\hphantom{.}\hfill IPPP/20/61\\
\hphantom{.}\hfill MSUHEP-20-017}
\author[a]{Bakul~Agarwal,}
\author[b,c]{Stephen~P.~Jones,}
\author[a]{and Andreas~von~Manteuffel}
\affiliation[a]{Department of Physics and Astronomy, Michigan State University,\\
East Lansing, Michigan 48824, USA}
\affiliation[b]{Institute for Particle Physics Phenomenology, Durham University, Durham DH1 3LE, UK}
\affiliation[c]{Theoretical Physics Department, CERN, 1211 Geneva 23, Switzerland}
\emailAdd{agarwalb@msu.edu}
\emailAdd{s.jones@cern.ch}
\emailAdd{vmante@msu.edu}
\abstract{
We calculate the two-loop QCD corrections to $gg \to ZZ$ involving a closed top-quark loop.
We present a new method to systematically construct linear combinations of Feynman integrals with a convergent parametric representation, where we also allow for irreducible numerators, higher powers of propagators, dimensionally shifted integrals, and subsector integrals.
The amplitude is expressed in terms of such finite integrals by employing syzygies derived with linear algebra and finite field techniques.
Evaluating the amplitude using numerical integration, we find agreement with previous expansions in asymptotic limits and provide ab initio results also for intermediate partonic energies and non-central scattering at higher energies.
}
\begin{document} 
\maketitle
\flushbottom
\section{Introduction}
\label{sec:introduction}
\noindent
$Z$ boson pair production is an essential process at the Large Hadron Collider (LHC).
Besides its immediate relevance as a signal process for precision physics~\cite{Aaboud:2017rwm,Aaboud:2019lxo,Aaboud:2019lgy,Sirunyan:2017zjc,Sirunyan:2020pub}, it is a significant background to on-shell and off-shell Higgs production for the four-lepton final state~\cite{Aaboud:2018puo,Sirunyan:2018sgc,Sirunyan:2019twz,ATLAS:2020wny}.
Continuum $Z$ pair production significantly contributes to off-shell Higgs production ($\sim10\%$) through interference effects \cite{Kauer:2012hd,Kauer:2013qba}. 
This is, in particular, important for indirect Higgs width constraints as proposed in \cite{Caola:2013yja,Campbell:2013una}.
The primary production channel for vector bosons at the LHC is quark-antiquark annihilation, which starts at tree level and is known to next-to-next-to-leading order (NNLO) QCD \cite{Cascioli:2014yka,Heinrich:2017bvg,Gehrmann:2014bfa,Caola:2014iua,Gehrmann:2015ora,Grazzini:2015hta,Kallweit:2018nyv}. 
The gluon fusion channel is loop-induced and starts formally at NNLO for the process $pp\to ZZ$. Nevertheless, it accounts for $O(60\%)$ \cite{Cascioli:2014yka} of the total NNLO correction owing to the high gluon luminosity at the LHC. 
Additionally, NLO corrections to $gg\to ZZ$ were also found to be quite sizable \cite{Caola:2015psa}, resulting in an $O(5\%)$ increase to the total $pp\to  ZZ$ cross section \cite{Grazzini:2018owa}.

The one-loop QCD amplitude for $gg\to ZZ$ was calculated a long time ago in \cite{Dicus:1987dj,Glover:1988rg,Zecher:1994kb}. At two-loops, the massless quark contribution was computed in \cite{vonManteuffel:2015msa,Caola:2015ila}.
It is expected that, due to the Goldstone boson equivalence theorem \cite{Lee:1977eg,Chanowitz:1985hj}, top-quark corrections at two-loops could be significant as well, especially for longitudinally polarised $Z$ bosons at high invariant mass.
This configuration is of particular interest, since it provides unique opportunities for measurement of an anomalous $t\overline{t}Z$ coupling \cite{Azatov:2016xik,Cao:2020npb}.
Contributions from top-quark at two-loops were calculated in \cite{Melnikov:2015laa,Caola:2016trd} using the large top-mass approximation and subsequently improved using Pad{\'e} approximants in \cite{Campbell:2016ivq}. In \cite{Grober:2019kuf}, an expansion around top-quark pair production threshold was incorporated with the large top-mass approximation for the form factors relevant for interference with the Higgs production amplitude, and in \cite{Davies:2020lpf}, the authors used both the large top-mass approximation and the small top-mass approximation along with Pad{\'e} approximants to improve the expansion in the intermediate region.
Higgs mediated two-loop contributions to $ZZ$ production involving a closed top-quark loop were calculated some time ago~\cite{Spira:1995rr,Harlander:2005rq,Anastasiou:2006hc,Aglietti:2006tp}.
Contributions of the third generation quarks to $W^+W^-$ production with exact mass dependence were computed recently in \cite{Bronnum-Hansen:2020mzk}.

In this paper, we calculate the two-loop QCD corrections to on-shell $gg\to ZZ$ production which involve a closed top-quark loop, keeping the dependence on the top-quark mass exact.
We present a new variant of the syzygy based approach for reduction of dimensionally regulated multi-loop integrals, which we use to reduce our amplitudes.
Since many of the topologies involved in this calculation are rather complicated and can not be expressed in terms of multiple polylogarithms, we use sector decomposition and numerically evaluate our master integrals.
To improve our numerical performance, we choose a basis of finite integrals, where we also allow for linear combinations of divergent integrals.
The building blocks of these linear combinations are rather general Feynman integrals, possibly with numerators, higher propagator powers (``dots''), pinched propagators (subsectors), or dimension shifts.
We present a new algorithm to systematically construct all possible linear combinations which are finite at the integrand level, starting from a set of seed integrals.

The paper is organised as follows. We introduce the setup for our amplitude calculation in section~\ref{sec:form_factors}, describing our projector method, the construction of helicity amplitudes and the electroweak coupling structure.
In section~\ref{sec:ibp}, we describe a new variant of the syzygy based approach to linear relations between loop integrals, which allows us to reduce the amplitude.
In section \ref{sec:finite_integrals}, we present our novel algorithm for construction of finite Feynman integrals, which we use to arrive at a basis of integrals suitable for numerical evaluation.
In section \ref{sec:checks}, we discuss UV renormalisation and IR subtraction, we then present the checks we perform on our calculation to establish correctness of our results.
Finally, we present numerical results for our helicity amplitudes in section \ref{sec:results}.
We detail some of our numerical checks in Appendix~\ref{sec:numchecks}.
\vfill
\section{Setup of the calculation}
\label{sec:form_factors}
\subsection{Form factors and helicity amplitudes}
We consider $Z$ pair production in gluon fusion,
\begin{equation}
\label{eq:2:1}
g(p_1)\,+\,g(p_2)\:\xrightarrow{}\:Z(p_3)\,+\,Z(p_4)\,.
\end{equation}
Here, $p_1,\,p_2$ are incoming and $p_3,\,p_4$ are outgoing momenta, so that $p_1 + p_2 = p_3 + p_4$ and
\begin{equation}
\label{eq:onshell}
p_1^2 = p_2^2 = 0 , \quad p_3^2 = p_4^2 = m_Z^2,
\end{equation}
that is, we consider the Z-bosons to be on-shell.
Our Mandelstam variables are
\begin{equation}
\label{eq:mandelstamvar}
s = (p_1+p_2)^2 \, , \quad t = (p_1-p_3)^2 \, , \quad u = (p_2-p_3)^2\,,
\quad\text{with~} s + t + u = 2\,m_Z^2\,.
\end{equation}
The amplitude can be represented as
\begin{equation}
\label{eq:amplitude}
\mathcal{M}=\mathcal{M}_{\mu\nu\rho\sigma}(p_1,p_2,p_3,p_4)\,\epsilon_{\lambda_1}^\mu(p_1)\,\epsilon_{\lambda_2}^\nu(p_2)\,\epsilon_{\lambda_3}^{*\rho}(p_3)\,\epsilon_{\lambda_4}^{*\sigma}(p_4)
\end{equation}
using polarization vectors $\epsilon_{\lambda_i}(p_i)$, for which we will also use the abbreviation $\epsilon_i\equiv\epsilon_{\lambda_i}(p_i)$.

Using Lorentz invariance, the amplitude can be decomposed in terms of 138 parity-even tensor structures~\cite{vonManteuffel:2015msa}:
\begin{align}
\label{eq:tensordecompose}
\mathcal{M}^{\mu\nu\rho\sigma}(p_1,p_2,p_3,p_4)\,&=\,a_1\,g^{\mu\nu}\,g^{\rho\sigma}\,+\,a_2\,g^{\mu\rho}\,g^{\nu\sigma}\,+\,a_3\,g^{\mu\sigma}\,g^{\nu\rho}\notag\\
&+\sum_{i,j=1}^3\,(\,a_{1,ij}\,g^{\mu\nu}\,p_i^\rho\,p_j^\sigma\,+\,a_{2,ij}\,g^{\mu\rho}\,p_i^\nu\,p_j^\sigma\,+\,a_{3,ij}\,g^{\mu\sigma}\,p_i^\nu\,p_j^\rho\notag\\
&\qquad\,\,+\,a_{4,ij}\,g^{\nu\rho}\,p_i^\mu\,p_j^\sigma\,+\,a_{5,ij}\,g^{\nu\sigma}\,p_i^\mu\,p_j^\rho\,+\,a_{6,ij}\,g^{\rho\sigma}\,p_i^\mu\,p_j^\nu\,)\notag\\
&+\sum_{i,j,k,l=1}^3\,a_{ijkl}\,p_i^\mu\,p_j^\nu\,p_k^\rho\,p_l^\sigma\,
\,.
\end{align}
Parity-odd tensor structures involving the epsilon tensor do not need to be taken into account due to Bose symmetry and charge-parity conservation for our process~\cite{Glover:1988rg}.
Since the color structure of the external states is straight-forward, we suppress color indices here and in the following. We can reduce the number of tensors using transversality of the gluon polarization vectors,
\begin{equation}
\label{eq:gluontransversality}
\epsilon_1\cdot p_1 = 0\,,\qquad\epsilon_2\cdot p_2 = 0\,,
\end{equation}
and the gauge choice
\begin{align}
\label{eq:externalbosongauge}
\epsilon_1 \cdot p_2 &= 0\,, &
\epsilon_2 \cdot p_1 &= 0\,, &
\epsilon_3 \cdot p_3 &= 0\,, &
\epsilon_4 \cdot p_4 &= 0\,.
\end{align}
These polarisation vectors correspond to the polarisation sums
\begin{align}
\label{eq:polsums}
\sum_{\text{pol}} \epsilon_1^\mu\,\epsilon_1^{*\nu} &= -g^{\mu\nu} + \frac{p_1^\mu p_2^\nu \,+\, p_2^\mu p_1^\nu}{p_1.p_2}\,, &
\sum_{\text{pol}} \epsilon_3^\mu\,\epsilon_3^{*\nu} &= -g^{\mu\nu} + \frac{p_3^\mu p_3^\nu}{p_3.p_3}\,,\notag\\
\sum_{\text{pol}} \epsilon_2^\mu\,\epsilon_2^{*\nu} &= -g^{\mu\nu} + \frac{p_1^\mu p_2^\nu \,+\, p_2^\mu p_1^\nu}{p_1.p_2}\,, &
\sum_{\text{pol}} \epsilon_4^\mu\,\epsilon_4^{*\nu}  &= -g^{\mu\nu} + \frac{p_4^\mu p_4^\nu}{p_4.p_4}\,.
\end{align}
The amplitude can then be written as
\begin{equation}
\label{eq:formfactors}
\begin{split}
\mathcal{M}^{\mu\nu\rho\sigma}(p_1,p_2,p_3,p_4)\,&=\,\sum_{i=1}^{20}\,A_i(s,t,m_t^2,m_Z^2)\,T_i^{\mu\nu\rho\sigma}\,,
\end{split}
\end{equation}
where the $A_i$ are the form factors, and the remaining 20 tensors $T_i$ are as follows:
\begin{align}
\label{eq:tensorsleft}
T^{\mu\nu\rho\sigma}_1 &= g^{\mu\nu}g^{\rho\sigma}\,, & 
T^{\mu\nu\rho\sigma}_2 &= g^{\mu\rho}g^{\nu\sigma}\,, & 
T^{\mu\nu\rho\sigma}_3 &= g^{\mu\sigma}g^{\nu\rho}\,, & 
T^{\mu\nu\rho\sigma}_4 &= p_1^{\rho}\,p_1^{\sigma}\,g^{\mu\nu}\,,\notag\\
T^{\mu\nu\rho\sigma}_5 &= p_1^{\rho}\,p_2^{\sigma}\,g^{\mu\nu}\,, & 
T^{\mu\nu\rho\sigma}_6 &= p_1^{\sigma}\,p_2^{\rho}\,g^{\mu\nu}\,, & 
T^{\mu\nu\rho\sigma}_7 &= p_2^{\rho}\,p_2^{\sigma}\,g^{\mu\nu}\,, & 
T^{\mu\nu\rho\sigma}_8 &= p_1^{\sigma}\,p_3^{\nu}\,g^{\mu\rho}\,,\notag\\
T^{\mu\nu\rho\sigma}_9 &= p_2^{\sigma}\,p_3^{\nu}\,g^{\mu\rho}\,, & 
T^{\mu\nu\rho\sigma}_{10} &= p_1^{\rho}\,p_3^{\nu}\,g^{\mu\sigma}\,, & T^{\mu\nu\rho\sigma}_{11} &= p_2^{\rho}\,p_3^{\nu}\,g^{\mu\sigma}\,, & T^{\mu\nu\rho\sigma}_{12} &= p_1^{\sigma}\,p_3^{\mu}\,g^{\nu\rho}\,,\notag\\
T^{\mu\nu\rho\sigma}_{13} &= p_2^{\sigma}\,p_3^{\mu}\,g^{\nu\rho}\,, & T^{\mu\nu\rho\sigma}_{14} &= p_1^{\rho}\,p_3^{\mu}\,g^{\nu\sigma}\,, & T^{\mu\nu\rho\sigma}_{15} &= p_2^{\rho}\,p_3^{\mu}\,g^{\nu\sigma}\,, & T^{\mu\nu\rho\sigma}_{16} &= p_3^{\mu}\,p_3^{\nu}\,g^{\rho\sigma}\,,\notag\\
T^{\mu\nu\rho\sigma}_{17} &= p_1^{\rho}\,p_1^{\sigma}\,p_3^{\mu}\,p_3^{\nu}\,, & 
T^{\mu\nu\rho\sigma}_{18} &= p_1^{\rho}\,p_2^{\sigma}\,p_3^{\mu}\,p_3^{\nu}\,, &
T^{\mu\nu\rho\sigma}_{19} &= p_2^{\rho}\,p_1^{\sigma}\,p_3^{\mu}\,p_3^{\nu}\,, &
T^{\mu\nu\rho\sigma}_{20} &= p_2^{\rho}\,p_2^{\sigma}\,p_3^{\mu}\,p_3^{\nu}\,.
\end{align}
The form factors $A_i(s,t,m_t^2,m_Z^2)$ can be derived from the amplitude using projection operators $P^{\mu\nu\rho\sigma}_i$, which fulfill
\begin{equation}
\label{eq:projectionoperators}
\sum_{\text{pol}} P_i^{\mu\nu\rho\sigma}
\epsilon^\ast_{1\mu}\epsilon^\ast_{2\nu}\epsilon_{3\rho}\epsilon_{4\sigma}
\epsilon_{1\mu'}\epsilon_{2\nu'}\epsilon^\ast_{3\rho'}\epsilon^\ast_{4\sigma'} \mathcal{M}^{\mu'\nu'\rho'\sigma'} = A_i\,.
\end{equation}
These projection operators themselves can be decomposed in terms of the $T_i^{\mu\nu\rho\sigma}$ as 
\begin{equation}
\label{eq:projopexpr}
P_i^{\mu\nu\rho\sigma}\,=\,\sum_{j=1}^{20}\,B_{ij}(s,t,m_t^2,m_Z^2)\,(T_j^{\mu\nu\rho\sigma})^\dagger\,, \qquad i=1,...,20\,.
\end{equation}
where the exact forms of the $B_{ij}$ are available at the \href{https://vvamp.hepforge.org/}{VVamp project website}. 

Due to Bose symmetry, the amplitude must remain unchanged under the exchange of the incoming gluons or the outgoing Z-bosons \cite{vonManteuffel:2015msa} i.e. 
\begin{equation}
\begin{split}
    1 \leftrightarrow 2 \qquad &: \quad p_1 \leftrightarrow p_2, \quad \epsilon_{\lambda_1}(p_1) \leftrightarrow \epsilon_{\lambda_2}(p_2),\notag\\
    3 \leftrightarrow 4 \qquad &: \quad p_3 \leftrightarrow p_4, \quad \epsilon_{\lambda_3}(p_3) \leftrightarrow \epsilon_{\lambda_4}(p_4).\notag
\end{split}
\end{equation}
This leads to the following identities between the form factors
\begin{equation}
\label{eq:ffidentities}
A_7 = A_4\,,\quad
A_{12} = -A_{11}\,,\quad
A_{13} = -A_{10}\,,\quad
A_{14} = -A_{9}\,,\quad
A_{15} = -A_8\,,\quad
A_{20} = A_{17}\,,
\end{equation}
as well as the following relations under the crossing $p_1 \leftrightarrow p_2$ ($t\leftrightarrow u$)
\begin{align}
\label{eq:ffcrossing}
A_{1}(s,t) &= A_{1}(s,u)\,, & A_{4}(s,t) &= A_{4}(s,u)\,, & A_{7}(s,t) &= A_{7}(s,u)\,,\notag\\
A_{16}(s,t) &= A_{16}(s,u)\,, & A_{17}(s,t) &= A_{17}(s,u)\,, & A_{20}(s,t) &= A_{20}(s,u)\,,\notag\\
A_2(s,t) &= A_3(s,u)\,, & A_5(s,t) &= A_6(s,u)\,, & A_8(s,t) &= A_{13}(s,u)\,,\notag\\
A_9(s,t) &= A_{12}(s,u)\,, & A_{10}(s,t) &= A_{15}(s,u)\,, & A_{11}(s,t) &= A_{14}(s,u)\,,\notag\\
A_{18}(s,t) &= A_{19}(s,u)\,. & & & & 
\end{align}

It is straightforward to derive amplitudes for polarised external particles from the $A_i$ form factors.
Taking fermionic decays of the $Z$ bosons into account, the amplitudes for specific fermion helicities can be found e.g.\ in \cite{vonManteuffel:2015msa}.
Here, we consider specific polarisations of the on-shell $Z$ bosons in the partonic center-of-mass frame.
We parametrise the momenta according to
\begin{align}
\label{eq:momentumchoice}
p_1^\mu &= \frac{\sqrt{s}}{2} \left(1,0,0,1 \right), &
p_3^\mu &= \frac{\sqrt{s}}{2} \left(1,\beta \sin\theta,0,\beta \cos\theta \right), \notag\\
p_2^\mu &= \frac{\sqrt{s}}{2} \left(1,0,0,-1 \right), &
p_4^\mu &= \frac{\sqrt{s}}{2} \left(1,-\beta \sin\theta,0,-\beta \cos\theta \right),
\end{align}
with $\beta=\sqrt{1-4m_Z^2/s}$ and $\theta$ being the angle in the centre-of-mass frame between the direction $p_1$ and the outgoing Z boson carrying momentum $p_3$.
We choose for the polarisation vectors
\begin{align}
\label{eq:helicities}
\epsilon^\mu_\pm(p_1) &= \frac{1}{\sqrt{2}}\left(0,\mp 1,-i,0\right),&& \notag\\
\epsilon^\mu_\pm(p_2) &= \frac{1}{\sqrt{2}}\left(0,\pm 1,-i,0\right),&& \notag\\
\epsilon^\mu_\pm(p_3) &= \frac{1}{\sqrt{2}}\left(0,\mp\cos\theta,-i,\pm\sin\theta \right),&
\epsilon^\mu_0(p_3) &= \frac{\sqrt{s}}{2 m_Z}\left(\beta,\sin\theta,0,\cos\theta\right),\notag\\ 
\epsilon^\mu_\pm(p_4) &= \frac{1}{\sqrt{2}}\left(0,\pm\cos\theta,-i,\mp\sin\theta \right),&
\epsilon^\mu_0(p_4) &= \frac{\sqrt{s}}{2 m_Z}\left(\beta,-\sin\theta,0,-\cos\theta\right).
\end{align}
It can be shown that these polarisation vectors satisfy \eqref{eq:gluontransversality}, \eqref{eq:externalbosongauge}, and \eqref{eq:polsums}. Moreover, the following symmetry relations hold for the helicity amplitudes \cite{Glover:1988rg,Davies:2020lpf}:
\begin{align}
\label{eq:helampsidentities}
    \quad\qquad\qquad \mathcal{M}_{\lambda_1 \lambda_2 \lambda_3 \lambda_4} &= (-1)^{\delta_{\lambda_3 0} + \delta_{\lambda_4 0}}\,\mathcal{M}_{-\lambda_1 -\lambda_2 -\lambda_3 -\lambda_4}\notag\\
    \mathcal{M}_{+++-} &= \mathcal{M}_{++-+}\notag\\
    \mathcal{M}_{+---} &= \mathcal{M}_{+-++}\notag\\
    \mathcal{M}_{++\pm 0} &= \mathcal{M}_{++0\pm}\notag\\
    \mathcal{M}_{+-\pm 0} &= \mathcal{M}_{+-0\mp}
\end{align}
as well as the following identities under the change $\theta\rightarrow\theta+\pi$:
\begin{align}
\label{eq:helampsrelations}
    \mathcal{M}_{++++} &= (\mathcal{M}_{++--})_{\theta\rightarrow\theta+\pi}\notag\\
    \mathcal{M}_{+-+-} &= (\mathcal{M}_{+--+})_{\theta\rightarrow\theta+\pi}\notag\\
    \mathcal{M}_{+\pm +0} &= (\mathcal{M}_{+\pm -0})_{\theta\rightarrow\theta+\pi}
\end{align}
which reduces the number of independent helicity amplitudes to 8. The expressions for the helicity amplitudes in terms of the form factors $A_i$ are provided in an ancillary file.

\subsection{Diagrams and electroweak coupling structure}
\label{sec:diags}
To generate the relevant Feynman diagrams, we use {\qgraf} \cite{NOGUEIRA1993279}. 
In the diagrams considered, the $Z$ bosons couple only to quark lines. The coupling of a $Z$ boson to a fermion line can be written as
\begin{align}
\label{eq:zbosonvertex}
    \mathcal{V}^{V f \bar{f}}_\mu &= ie\left[ L^Z_{f\bar{f}} \gamma_\mu \left( \frac{1-\gamma_5}{2} \right) + R^Z_{f\bar{f}} \gamma_\mu \left( \frac{1+\gamma_5}{2} \right) \right]\notag\\
    &= i\frac{e}{2 \sin\theta_W\cos\theta_W} \gamma_\mu \left( v_t + a_t \gamma_5 \right)
\end{align}
where $L^Z_{f\bar{f}} = (I_3^f - q_f \sin^2\theta_W)/(\sin \theta_W \cos \theta_W)$, $R^Z_{f\bar{f}} = - q_f \sin\theta_W/\cos \theta_W$, $e$ is the positron charge, and $q_f$ is the electric charge of the fermion in terms of $e$.
The vector and axial components are given in terms of the weak mixing angle $\theta_W$ by $v_t = \frac{1}{2} - \frac{4}{3} \sin^2 \theta_W$ and $a_t = -\frac{1}{2}$, respectively.
The couplings of the two Z bosons to the fermion line can in principle generate vector-vector ($v_t^2$), vector-axial ($v_t a_t$), and axial-axial ($a_t^2$) contributions to the amplitude. However, due to Bose symmetry and charge-parity conservation for this process, the vector-axial contribution should vanish identically \cite{Glover:1988rg}. This also explains the absence of any terms with the Levi-Civita tensor in \eqref{eq:tensordecompose} since such terms would violate parity and are hence forbidden. For the massless quark case, the vector-vector and the axial-axial contributions are identical; after including quark masses, they differ by terms proportional to the quark mass.

\begin{figure}
\centering
\raisebox{12.5mm}{\textbf{[A]}}
\includegraphics[height=25mm]{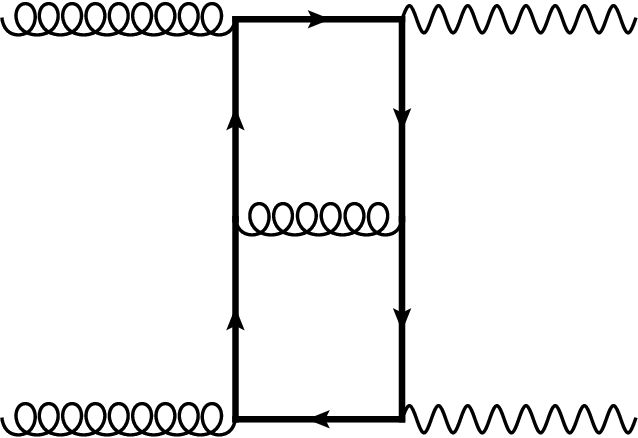}
\qquad\qquad
\raisebox{12.5mm}{\textbf{[B]}}
\includegraphics[height=25mm]{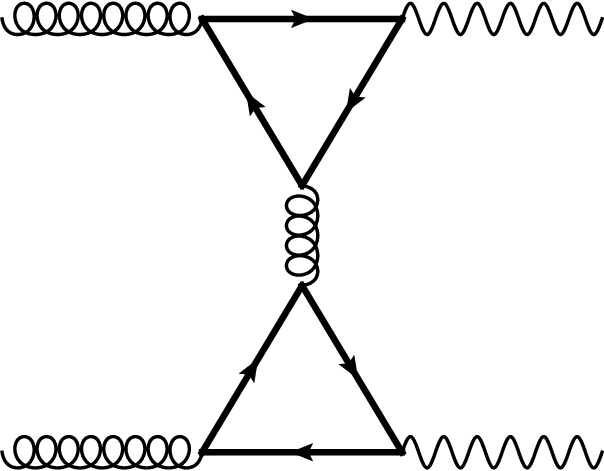}
\caption{Example Feynman diagrams representing the two classes of diagrams}
\label{fig:classes}
\end{figure}

We find a total of 166 diagrams containing at least one top-quark propagator. Out of these, 49 diagrams have a single gluon coupled to a closed fermion loop, and hence they vanish due to colour conservation. The remaining diagrams can be divided into two classes shown in figure \ref{fig:classes}.

\textbf{Class A} : Both Z bosons couple to the same fermion line. To appropriately handle $\gamma_5$ in $d$ dimensions, we use the anti-commuting $\gamma_5$ scheme described in \cite{Kreimer:1989ke,Korner:1991sx}. Since cyclicity of trace is not preserved in this scheme, a reading point prescription is employed to ensure that all traces are read from the same point. However, for a closed fermion loop with an even number of $\gamma_5$ matrices, it is trivial to eliminate $\gamma_5$ using the anti-commutation relations; this greatly simplifies the implementation of the anti-commuting $\gamma_5$ scheme.

\textbf{Class B} : The Z bosons couple to different closed fermion lines. For these diagrams, the vector-vector contribution can be shown to vanish due to Furry's theorem, while the vector-axial piece is identically zero because of charge-parity conservation. The axial-axial piece, however, vanishes only after summing over a degenerate $SU(2)_L$ doublet. Since the third generation of quarks is not degenerate, this cancellation is incomplete and we see a finite remainder from the top-bottom mass splitting. These diagrams have a single $\gamma_5$ in each loop which leads to a non-trivial structure and requires careful application of the reading point prescription. Since these diagrams are effectively one-loop, we treat them separately. Exact results for these diagrams were previously presented in \cite{Campbell:2016ivq} and we find full agreement. We will not mention them any further and do not include these contributions in the results presented below.

\begin{figure}
\begin{subfigure}[c]{0.22\textwidth}
\includegraphics[width=0.8\linewidth, center]{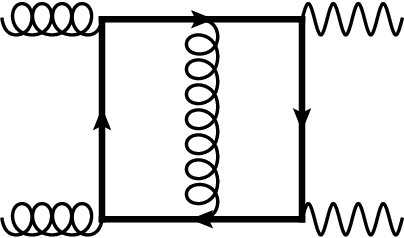}
\caption{}
\label{fig:irredtop:1}
\end{subfigure}
\begin{subfigure}[c]{0.22\textwidth}
\includegraphics[width=0.8\linewidth, center]{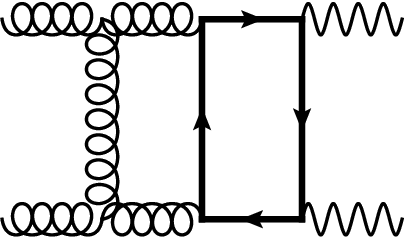}
\caption{}
\label{fig:irredtop:3}
\end{subfigure}
\begin{subfigure}[c]{0.22\textwidth}
\includegraphics[width=0.8\linewidth, center]{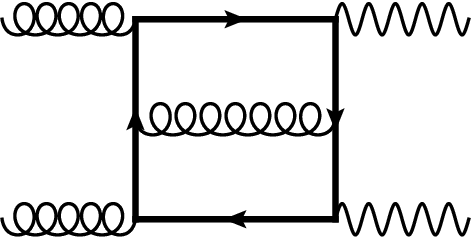}
\caption{}
\label{fig:irredtop:2}
\end{subfigure}
\begin{subfigure}[c]{0.22\textwidth}
\includegraphics[width=0.8\linewidth, center]{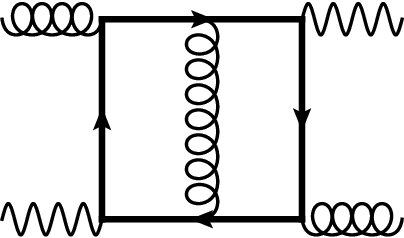}
\caption{}
\label{fig:irredtop:4}
\end{subfigure}
\begin{subfigure}[c]{0.22\textwidth}
\includegraphics[width=0.8\linewidth, center]{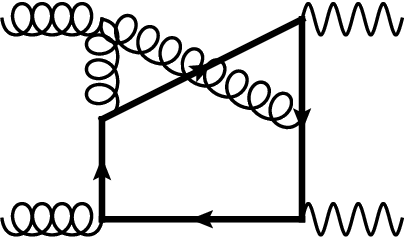}
\caption{}
\label{fig:irredtop:5}
\end{subfigure}
\begin{subfigure}[c]{0.22\textwidth}
\includegraphics[width=0.8\linewidth, center]{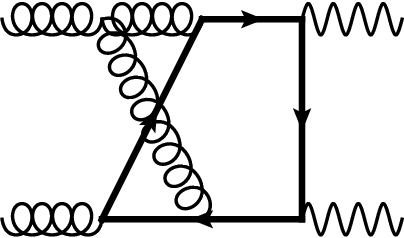}
\caption{}
\label{fig:irredtop:6}
\end{subfigure}
\centering
\begin{subfigure}[c]{0.22\textwidth}
\includegraphics[width=0.8\linewidth, center]{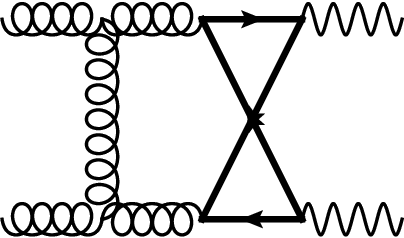}
\caption{}
\label{fig:irredtop:7}
\end{subfigure}
\caption{Representative Feynman diagrams in class A with irreducible topologies. The number of master integrals in each topology are 3, 4, 3, 3, 5, 5, and 4 respectively}
\label{fig:irredtop}
\end{figure}

\begin{figure}
\centering
\begin{subfigure}[c]{0.22\textwidth}
\includegraphics[width=0.9\linewidth, center]{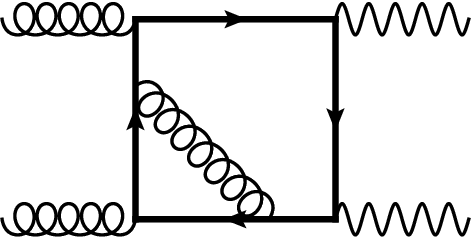}
\caption{}
\label{fig:redtop:1}
\end{subfigure}
\begin{subfigure}[c]{0.22\textwidth}
\includegraphics[width=0.9\linewidth, center]{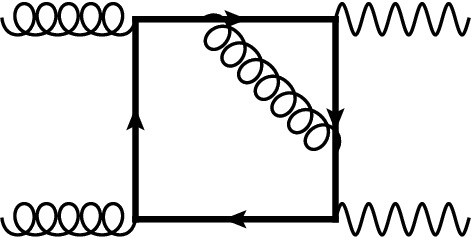}
\caption{}
\label{fig:redtop:2}
\end{subfigure}
\begin{subfigure}[c]{0.22\textwidth}
\includegraphics[width=0.9\linewidth, center]{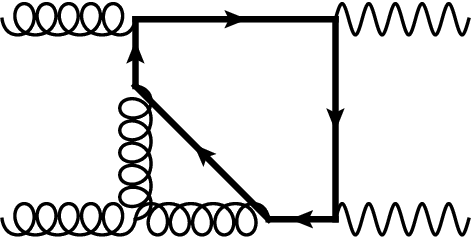}
\caption{}
\label{fig:redtop:3}
\end{subfigure}
\\
\begin{subfigure}[c]{0.22\textwidth}
\includegraphics[width=0.9\linewidth, center]{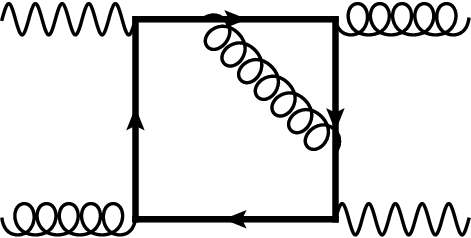}
\caption{}
\label{fig:redtop:4}
\end{subfigure}
\begin{subfigure}[c]{0.22\textwidth}
\includegraphics[width=0.9\linewidth, center]{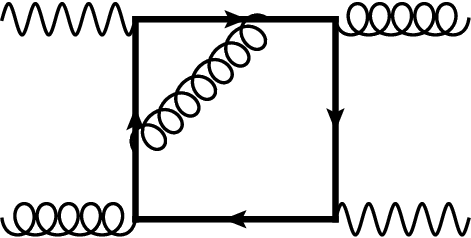}
\caption{}
\label{fig:redtop:5}
\end{subfigure}
\begin{subfigure}[c]{0.22\textwidth}
\includegraphics[width=0.9\linewidth, center]{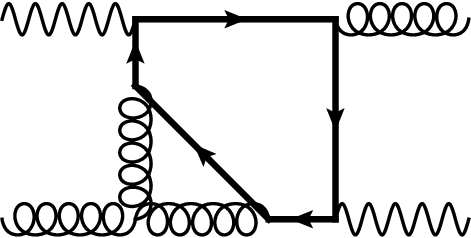}
\caption{}
\label{fig:redtop:6}
\end{subfigure}
\caption{Representative Feynman diagrams in class A with reducible topologies.}
\label{fig:redtop}
\end{figure}

After generating the diagrams in class A, we employ {\reduze} to map them to the 4 different integral families shown in table \ref{tab:families}.
 We find 13 top-level topologies (trivalent graphs), of which 7 are irreducible (figure \ref{fig:irredtop}) and 6 are reducible (figure \ref{fig:redtop}).
 We use {\form} \cite{Ruijl:2017dtg,Kuipers:2012rf,Vermaseren:2000nd} to apply the Feynman rules and generate the amplitude, employing the 't Hooft-Feynman gauge $(\xi=1)$ for internal gluons. Before applying any symmetries, we find a total of 29247 integrals with up to 4 irreducible scalar products in the numerator. This number can be reduced to 4504 using symmetry relations between the integrals. We see further simplifications after inserting the symmetry relations in the amplitude; due to cancellations only 1584 integrals survive in the form factors. This is a significant improvement over the original number of integrals and underlines the importance of using symmetry relations and working with amplitudes instead of individual diagrams.

\begin{table}
{\small
\centering
\begin{tabular}{ |c|c|c|c| } 
 \hline
 A & B & C & D \\ 
 \hline\hline
 $k_1^{\,2}-m_t^{\,2}$ & $k_1^{\,2}$ & $k_1^{\,2}-m_t^{\,2}$ & $k_1^{\,2}$ \\
 \hline
 $(k_1+p_1)^{\,2}-m_t^{\,2}$ & $(k_1+p_1)^{\,2}$ & $(k_1+p_1)^{\,2}-m_t^{\,2}$ & $(k_1+p_1)^{\,2}$ \\
 \hline
 $(k_1+p_1+p_2)^{\,2}-m_t^{\,2}$ & $(k_1+p_1+p_2)^{\,2}$ & $(k_1+p_1-p_3)^{\,2}-m_t^{\,2}$ & $(k_1+p_1+p_2)^{\,2}$ \\
 \hline
 $(k_1+p_4)^{\,2}-m_t^{\,2}$ & $(k_1+p_4)^{\,2}$ & $(k_1+p_4)^{\,2}-m_t^{\,2}$ & $k_2^{\,2}-m_t^{\,2}$ \\
 \hline
 $k_2^{\,2}-m_t^{\,2}$ & $k_2^{\,2}-m_t^{\,2}$ & $k_2^{\,2}-m_t^{\,2}$ & $(k_2+p_1+p_2)^{\,2}-m_t^{\,2}$ \\
 \hline
 $(k_2+p_1)^{\,2}-m_t^{\,2}$ & $(k_2+p_1-p_3)^{\,2}-m_t^{\,2}$ & $(k_2+p_1-p_3)^{\,2}-m_t^{\,2}$ & $(k_2+p_4)^{\,2}-m_t^{\,2}$ \\
 \hline
 $(k_2+p_1+p_2)^{\,2}-m_t^{\,2}$ & $(k_2-p_3)^{\,2}-m_t^{\,2}$ & $(k_2+p_4)^{\,2}-m_t^{\,2}$ & $(k_2-k_1)^{\,2}-m_t^{\,2}$ \\
 \hline
 $(k_2+p_4)^{\,2}-m_t^{\,2}$ & $(k_2+p_4)^{\,2}-m_t^{\,2}$ & $(k_2-k_1)^{\,2}$ & $(k_2-k_1+p_2)^{\,2}-m_t^{\,2}$ \\
 \hline
 $(k_1-k_2)^{\,2}$ & $(k_1+k_2+p_4)^{\,2}-m_t^{\,2}$ & $(k_1-k_2+p_1)^{\,2}$ & $(k_2-k_1+p_4)^{\,2}-m_t^{\,2}$ \\
 \hline
\end{tabular}
}
\caption{List of integral families and their propagators}
\label{tab:families}
\end{table}

Using {\reduze} \cite{vonManteuffel:2012np,Studerus:2009ye,Bauer:2000cp,fermat}, we perform a numerical reduction by substituting numbers for kinematics and find, for the diagrams with a single fermion loop, 85 irreducible topologies with the worst sector having 6 master integrals with 6 lines. In total, we obtain 264 master integrals for class A, out of which 172 are not related by any crossing.
Our symbolic reduction is discussed in the following section.

\section{Reduction of Feynman integrals}
\label{sec:ibp}
\subsection{Linear relations from syzygies}
A general $L$-loop scalar Feynman integral with $N$ propagators can be represented by
\begin{equation}
\label{eq:feynnmanintegral}
    I(\nu_1,...,\nu_N) = \int \left(\prod^{L}_{l=1} \ud^d k_l\right) \,\prod^{N}_{i=1}\,\frac{1}{{(q_i^{2}-m_i^{2})}^{\nu_i}}
\end{equation}
where $k_1,...,k_L$ are the loop momenta, $q_i$ are the propogator momenta (linear combinations of loop and external momenta), $m_i$ are the masses of the propagators, $\nu_i$ are (integer) exponents of the propagators, and $d=4-2\epsilon$.
Here, we allow also for non-positive powers $\nu_i$ of the propagators, i.e.\ we consider a \emph{family} of integrals with possible irreducible numerators.
The total derivative of an integral in dimensional regularisation vanishes; this allows us to write linear relations between different integrals \cite{Chetyrkin:1981qh}
\begin{equation}
\label{eq:momspaceibp}
    0 = \int \left(\prod^{L}_{l=1} \ud^d k_l\right)\,\frac{\partial}{\partial k_j^\mu}\left(v^\mu\,\prod^{N}_{i=1}\,\frac{1}{{(q_i^{2}-m_i^{2})}^{\nu_i}}\right),
\end{equation}
where $v_\mu$ could be any linear combination of loop and external momenta. We can eliminate most of the integrals in the amplitude using these relations with the remaining integrals usually referred to as master or basis integrals. This procedure can be systematically used to reduce any integral appearing in the amplitude due to an algorithm by S.\ Laporta \cite{Laporta:2001dd}. Many public codes based on this algorithm are available for this purpose \cite{Anastasiou:2004vj,Lee:2012cn,vonManteuffel:2012np,Maierhoefer:2017hyi,Smirnov:2019qkx}.

Conventionally, the vector $v_\mu$ is chosen as a single loop or external momentum; different such choices yield a set of simple equations as a starting point. It is easy to see that the derivatives in \eqref{eq:momspaceibp} generate higher powers $\nu_i$ of the propagators, often referred to as ``dots''. Such auxiliary integrals with a large number of dots are usually not required for the amplitude and lead to relatively large linear systems that are computationally expensive to reduce. 

A method was proposed in \cite{Gluza:2010ws} to avoid these higher powers of propagators by constructing suitable generating vectors $v_\mu$ from syzygies. This method involves the computation of a Gr\"obner basis to obtain the syzygies. A linear algebra based approach was presented in \cite{Schabinger:2011dz}, albeit the syzygies can only be obtained to a specified degree using this method.
Subsequent work \cite{Lee:2013mka,Ita:2015tya,Larsen:2015ped} refined syzygy based constructions in the momentum space representation as well as in Baikov's representation~\cite{Baikov:1996rk}.
Syzygies can also be used to derive linear relations \cite{Lee:2014tja,Bitoun:2017nre,vonManteuffel:2020vjv} in the Lee-Pomeransky representation~\cite{Lee:2013hzt}.

The $L$-loop Feynman integral in \eqref{eq:feynnmanintegral} can be written in Baikov's representation as
\begin{equation}
\label{eq:feynmanintbaikov}
    I(\nu_1,...,\nu_N) = \mathcal{N}_0 \int \ud z_1...\ud z_N\,\,\frac{1}{\prod^{N}_{i=1}\,z_i^{\nu_i}}\,\,P^{\frac{d-L-E-1}{2}}\,,
\end{equation}
where the Jacobian of the variable transformation involves the determinant $P$, the \emph{Baikov polynomial}, $\mathcal{N}_0$ is a normalization factor,
and $E$ is the number of linearly independent external momenta. The integration-by-parts identities in Baikov's representation are given by
\begin{equation}
\label{eq:baikovibp}
    0 = \int \ud z_1...\ud z_N\,\,\sum^N_{i=1}\,\left(\,\frac{\partial{f_i}}{\partial z_i}\,+\frac{d-L-E-1}{2P}\,f_i\,\frac{\partial{P}}{\partial{z_i}}\,-\nu_i\,\frac{f_i}{z_i}\right)\,\frac{1}{\prod^{N}_{i=1}\,z_i^{\nu_i}}\,\,P^{\frac{d-L-E-1}{2}}\,,
\end{equation}
where $f_1$, \ldots, $f_N$ are arbitrary polynomials in the Baikov parameters $z_1$, \ldots, $z_N$, and the kinematic invariants.
In the above equation, terms that appear with $1/P$ lead to dimensionally shifted integrals. Since these integrals don't appear in the amplitude, it may be desirable to avoid them to prevent an unnecessary proliferation of auxiliary quantities in the system. This can be achieved by imposing the constraint
\begin{equation}
\label{eq:nodimsyzygy}
    \left(\,\sum^N_{i=1}\,f_i\,\frac{\partial{P}}{\partial{z_i}}\,\right)\,+\,f_{N+1}\,P = 0\,.
\end{equation}
Here, we introduced a new polynomial $f_{N+1}$ in the Baikov parameters. Note that $P$ and its derivatives are known polynomials for the problem, see e.g.\ \cite{Boehm:2017wjc} for details.
A constraint of this type on the vector of polynomials $(f_1,\ldots,f_{N+1})$ is known as a syzygy in algebraic geometry. Explicit solutions to this equation were pointed out in \cite{Boehm:2017wjc} and can easily be written down.
The resulting $f_i$ are linear polynomials in the Baikov variables $z_k$ and the kinematic invariants. It must be noted that these $f_i$ generate integration-by-parts relations which cover~\cite{Boehm:2017wjc} those derived in the conventional momentum-space approach \eqref{eq:momspaceibp}.

To enforce the absence of doubled propagators, one requires that for all $i$ with $\nu_i\ge 1$, the $f_i$ are proportional to $z_i$ to cancel the $1/z_i$ in the relation,
\begin{equation}
\label{eq:nodotsyzygy}
    f_i\,=\,b_i\,z_i\quad \forall\,i=1,\ldots,N~\text{with~}\nu_i \geq 1\,.
\end{equation}
While it is straight-forward to fulfil both constraints \eqref{eq:nodimsyzygy} and \eqref{eq:nodotsyzygy} separately, a simultaneous solution requires a non-trivial calculation.

\subsection{Constructing syzygies with linear algebra}
Formally, finding vectors of polynomials $(f_i)$ which are simultaneous solutions of both \eqref{eq:nodimsyzygy} and \eqref{eq:nodotsyzygy} corresponds to the determination of the intersection of two syzygy modules \cite{Boehm:2018fpv}.
In practice, computer algebra packages implement algorithms to solve this task.
For performance reasons we decided to develop a custom syzygy solver based on linear algebra and finite field arithmetic \cite{vonManteuffel:2014ixa,Peraro:2016wsq}. Note that if polynomials $(f_i)$ satisfy the syzygy constraint in \eqref{eq:nodimsyzygy}, then $(z_k f_i)$ for any $k$ also satisfy it.

\begin{algorithm}
\caption{Syzygies for linear relations without dimension shifts or dots}
\emph{Input:} Syzygies of degree 1 solving \eqref{eq:nodimsyzygy}, maximal required degree $n_\text{max}$.\\
\emph{Output:} Syzygies $S_1,\ldots,S_{n_\text{max}}$ up to degree $n_\text{max}$ solving \eqref{eq:nodimsyzygy} and \eqref{eq:nodotsyzygy}.

\begin{algorithmic}[1]
\State Start with syzygies of degree $n=1$. Let $I_1$ be a complete set of solutions $(f_i)$ to the no-dimension-shift constraint~\eqref{eq:nodimsyzygy}, which are linear in the Baikov parameters $z_k$. These can directly be written down \cite{Boehm:2017wjc}.
Abbreviating the momenta squared with variables $z_{N+1}$, \ldots, the vectors in $I_1$ are of homogeneous degree 1 in the variables $z_k$.

\State At degree $n$, form a matrix $M_n$, where each element of $(f_i)\in I_n$ corresponds to a row.
The columns enumerate both the component $i$ of $(f_i)$ and the power products of $z_i$ in them; the entries of the matrix are the coefficients.
A column is called admissable, if it satisfies the no-doubled propagator constraint \eqref{eq:nodotsyzygy}, and non-admissable otherwise.
All admissable columns are ordered to the right of the non-admissable columns.

\State Perform a row reduction of $M_n$. In the row reduced form, select all rows, which have an admissable pivot column and form the corresponding syzygies $S_n$ from them.
$S_n$ forms a complete set of linear combinations of the syzygies in $I_n$, which satisfy \eqref{eq:nodotsyzygy} for all of their terms, and are therefore our solutions at degree $n$.

\State If $n$ is the user-defined maximal degree, stop and return the solutions $S_1$, \ldots, $S_n$. Otherwise, proceed.

\State For each vector of polynomials $(f_i)\in I_n$ and each $z_k$, form the vector of polynomials $(z_k f_i)$. This gives the set $I_{n+1}$, which are solutions of \eqref{eq:nodimsyzygy} of degree $n+1$ in the $z_k$ but not necessarily solutions of \eqref{eq:nodotsyzygy}.

\State Replace $n \to n+1$ and go to step 2.
\end{algorithmic}
\label{alg:syz}
\end{algorithm}

In algorithm~\ref{alg:syz}, we provide a description of our method which converts
the intersection problem up to a specific degree of the syzygies to row reduction of a matrix.
Here, we treat the kinematic invariants as indeterminates of the polynomial ring, such that the matrices $M_n$ have entries which are rational numbers.
Alternatively, one can treat the invariants as part of the coefficient field.
This decreases the number of columns of the matrices $M_n$, but the entries are then rational functions of the kinematic invariants.
Since in the second approach the kinematic invariants do not count towards the degree $n$ in our algorithm, a lower maximal value may be sufficient for the integral reduction problem at hand compared to the first approach.
It is useful to use e.g.\ an overall mass dimension squared as a homogenizing variable $z_{N+1}$ for the last component of the syzygy vectors in this setup.

The row reduction of the matrix $M_n$ eliminates redundancies between the syzygies at degree $n$.
In our approach, we generate templates for the generation of linear relations between Feynman integrals from the syzygies.
We allow the templates to be applied to seed integrals with specific integer propagator powers and perform a subsequent row reduction on the resulting identities, similar to the traditional Laporta algorithm.
In this approach, we find it useful to filter out syzygies that are just a lower degree syzygy multiplied with an overall power product in the $z_k$.
This is achieved by determining reducible monomials using the row reduced form of an auxiliary matrix for the syzygies induced by lower degree syzygies~\cite{CabarcasDing}.

For our current process, we generated the required syzygies and performed the subsequent Laporta step with an in-house linear solver, Finred, based on finite field arithmetic and rational reconstruction. To simplify the linear relations further, we set $m_t=1$ and use a numerical value for the Z-boson mass as a ratio over top-quark mass, $m_Z^2/m_t^2=5/18$. This amounts to factoring out powers of $m_t^2$ corresponding to the mass dimension of the respective form factor.
In this way, we successfully reduced all of the Feynman integrals in our calculation to master integrals. The reductions proved to be rather challenging nevertheless and required significant computational resources. This is evident from the fact that the reduction tables exceeded 200 GB in size, with rational functions of degrees of up to 190 in the kinematic variables appearing in the reduction tables. The non-planar topologies, unsurprisingly, were the most difficult and accounted for almost all of the computation time and disk space. An interesting point to note is that within the planar topologies, figures \ref{fig:irredtop:1}, \ref{fig:irredtop:2}, and \ref{fig:irredtop:3}, with adjacent gluons are significantly simpler than \ref{fig:irredtop:4} with the gluons at the opposite vertices.

\subsection{Inserting the reductions into the amplitude}
\label{sec:backsub}
After having generated the reduction identities, the next task is to insert them into the unreduced amplitude. The reduction identities for this process are very complicated with a size of over 200 GB. As such, this task in itself is a major challenge. We used several tools and techniques to make this more manageable. 

We first calculate the reduction identities to the conventional Laporta basis and perform multivariate partial-fractioning of the reduction tables based on polynomial reductions with respect to a Gr\"obner basis~\cite{Abreu:2018zmy,MultivariateApart}. We implement this using the public code {\singular}~\cite{DGPS}~\footnote{Recently, an alternative method was developed in~\cite{Boehm:2020ijp}} with a polynomial ordering that prefers polynomials with lower degrees in kinematic variables and smaller coefficients, and are able to drastically reduce the size of the reduction tables.
We found it useful to first perform the partial fractioning for the $d$ dependent denominators and then partial fraction the kinematic denominators.
We note that this procedure can also be used even in the presence of denominators which depend both on $d$ and the kinematic variables.

We use custom {\form} scripts to insert the reduction identities into the amplitudes, and again perform multivariate partial-fractioning on the reduced amplitudes to arrive at a simpler representation in terms of kinematic variables and the irreducible denominators.
We see a drastic level of compression at this step; after partial-fractioning, the total size of amplitudes reduces from $\sim$300 GB to $\sim$600 MB.

Next, we perform a change of basis to express our amplitudes in terms of finite integrals.
We explain this choice of basis in more detail in section~\ref{sec:finite_integrals}.
After partial fractioning the basis change identities, we insert them into the reduced amplitude to arrive at the final reduced amplitude in terms of our finite basis. This step was computationally expensive, with more than a week of run-time and intermediate expressions with sizes in the terabytes before partial fractioning.
Once we have the amplitudes in the new choice of basis integrals, we perform partial-fractioning to simplify them.
Note that the form factors in the conventional Laporta basis contain many denominator factors that are polynomials in both $d$ and kinematics; we find that all such denominators no longer appear for our choice of finite master integrals.

As a last simplification measure we expand the form factors around $d=4$.
Our projectors introduce a spurious pole of order $1/\epsilon^5$, which cancels after reduction.
Since we are calculating an NLO amplitude, UV and IR subtractions will involve at most $1/\epsilon$ and $1/\epsilon^2$ poles, respectively.
The reduced bare form factors should therefore not have any pole worse than $1/\epsilon^2$, which, however, is not completely manifest when using our symbolic master integrals.
However, the change of basis to finite integrals removes the $1/\epsilon^4$ poles at the algebraic level.
In section~\ref{sec:checks} we describe in detail how all the poles show the expected behaviour with high numerical precision.

In the final representation, we are able to bring down the size of the worst coefficients to less than $1\,\mathrm{MB}$.
We create a {\tt C++} library for fast evaluation of the integral coefficients, either with exact rational arithmetic or with arbitrary precision floating point arithmetic using the {\tt GMP} library.
Even though the expressions are still sizable, we can evaluate all coefficients for a generic point in phase space within half a minute using rational arithmetic or within $3 s$ using floating point arithmetic with a target precision of 15 digits on a single CPU core.

\section{Finite basis integrals}
\label{sec:finite_integrals}
\subsection{Dimension shifts and dots}
To evaluate the master integrals, a powerful approach is to use differential equations to find analytic solutions \cite{Kotikov:1990kg,Remiddi:1997ny,Gehrmann:1999as,Argeri:2007up,Henn:2013pwa,Heller:2019gkq}.
This approach was used to calculate the master integrals in terms of multiple polylogarithms for the 2-loop massless corrections to diboson production in \cite{Gehrmann:2014bfa,Gehrmann:2015ora,Henn:2014lfa,Caola:2014lpa}.
Due to the massive top-quark loop in the corrections considered here, we expect the presence of functions beyond multiple polylogarithms, which makes the evaluations of master integrals considerably more challenging. 
While there has been significant progress concerning the analytic evaluation of Feynman integrals beyond polylogarithms \cite{Bonciani:2016qxi,vonManteuffel:2017hms,Broedel:2018iwv,Adams:2018kez,Lee:2018jsw,Walden:2020odh}, integrals of the type considered here remain a challenge.
An alternative is the use of expansions to solve the differential equations numerically~\cite{Aglietti:2007as,Lee:2017qql,Francesco:2019yqt,Hidding:2020ytt,Bronnum-Hansen:2020mzk}.
Here, we use a purely numerical approach to integrate the master integrals, namely sector decomposition \cite{Binoth:2000ps,Bogner:2007cr,Borowka:2018goh,Smirnov:2015mct}; see also \cite{Borowka:2016ypz,Chen:2020gae} for recent applications.

A naive integration-by-parts reduction using Laporta's algorithm with a generic ordering criterion leads to a conventional basis of master integrals. This basis is rather difficult to evaluate numerically since the integrals are often divergent and numerically unstable, and as such is inadequate for our purpose. We instead choose a different basis of master integrals that is finite in the limit $d\rightarrow4$. It was observed in \cite{Borowka:2016ypz,vonManteuffel:2017myy} that using a basis of finite integrals is highly beneficial, leading to a numerically more stable behaviour. Additionally, finite integrals often require fewer orders in the $\epsilon$ expansion which, coupled with better numerical stability, improves the overall performance significantly.

\begin{figure}[t]
\centering
\begin{subfigure}[c]{0.45\textwidth}
\raisebox{-.5\height}{\includegraphics[width=0.5\linewidth, center]{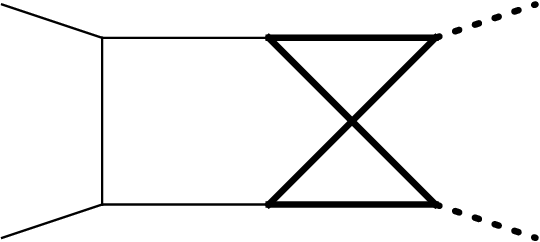}}
\caption{Divergent integral in $d=4-2\epsilon$}
\label{fig:dimshiftints:1}
\end{subfigure}
\begin{subfigure}[c]{0.45\textwidth}
$(k^2-m_t^2)$\hspace{-1.5cm}\raisebox{-.5\height}{\includegraphics[width=0.5\linewidth, center]{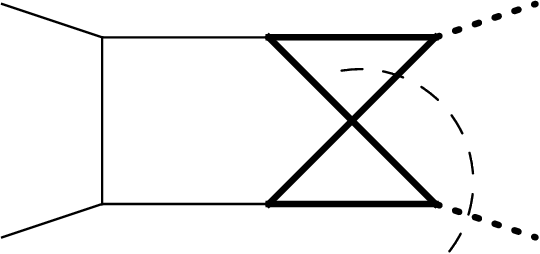}}
\caption{Divergent integral in $d=4-2\epsilon$ with an irreducible numerator}
\label{fig:dimshiftints:2}
\end{subfigure}
\begin{subfigure}[c]{0.45\textwidth}
\raisebox{-.5\height}{\includegraphics[width=0.58\linewidth, center]{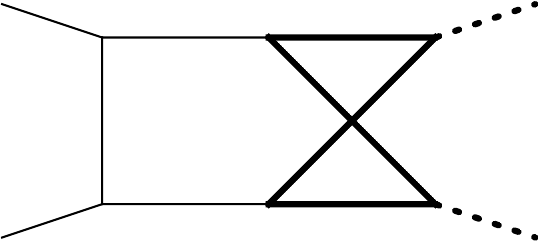}}
\caption{Finite integral in $d=6-2\epsilon$}
\label{fig:dimshiftints:3}
\end{subfigure}
\begin{subfigure}[c]{0.45\textwidth}
\raisebox{-.5\height}{\includegraphics[width=0.5\linewidth, center]{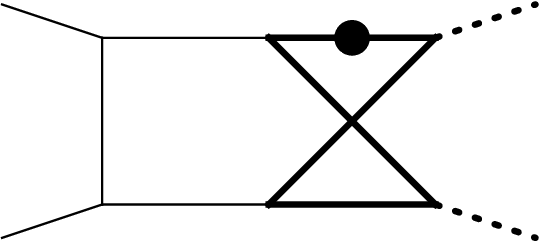}}
\caption{Finite integral with a dot in $d=6-2\epsilon$}
\label{fig:dimshiftints:4}
\end{subfigure}
\caption{Examples of divergent and finite integrals in the limit $\epsilon \to 0$ for a non-planar topology.
Thick solid lines represent the top-quark while thick dashed lines represent Z-bosons.
Topology (b) contains an irreducible numerator, where $k$ is the difference of the momenta of the edges marked by the thin dash lines.}
\label{fig:dimshiftints}
\end{figure}

One possible approach to constructing finite integrals is to use dimensionally shifted integrals \cite{Bern:1992nf}, possibly with doubled (or higher powers of) propagators.
It is always possible to construct a basis in this way \cite{Panzer:2014gra,vonManteuffel:2014qoa} and also straightforward in practice using e.g.\ the finite integral finder in {\reduze}.
Examples of such integrals are shown in figure \ref{fig:dimshiftints}. While it is convenient to find such finite integrals with dimension shifts and dots, they require computation of additional reduction identities beyond those required for the amplitude. For example, reductions for integrals with 2 additional dots are required for the dimension shift of two-loop integrals, and typically such integrals do not directly appear in the amplitude calculation.
It may therefore seem interesting to consider alternative choices of finite integrals.

Here, we explore a different approach by constructing finite integrals through linear combinations of divergent integrals based on the Feynman parametric representation~\cite{Agarwal:2019rag}.
In such linear combinations, non-integrable divergences of individual integrals cancel at the integrand level. This results in a single generalised Feynman parameter integral that is finite. We briefly describe the algorithm in the following subsection.

\subsection{Constructing finite linear combinations}
Consider a general $L$-loop integral in $d$ dimensions with $N$ distinct propagators in the momentum space representation,
\begin{equation}
\label{eq:4:1}
    I(\nu_1,...,\nu_N) = \int \left(\prod^{L}_{l=1}\frac{\ud^d k_l}{i\pi^{d/2}}\right)\,\prod^{N}_{j=1}\,\frac{1}{{(q_j^{2}-m_j^{2}+i\,\epsilon)}^{\nu_j}}
\end{equation}
with integer exponents $\nu_j\in\mathbb{Z}$.
If all indices $\nu_j$ are positive, one can use (see e.g.\ \cite{Smirnov:2004ym,Heinrich:2020ybq})
\begin{align}
\label{eq:line}
    \frac{1}{{(q_j^{2}-m_j^{2}+i\,\epsilon)}^{\nu_j}}
    &= \frac{(-1)^{\nu_j}}{\Gamma(\nu_j)} \int_0^\infty \ud x_j\,x_j^{\nu_j-1}\, e^{x_j\,(q_j^{2}-m_j^{2}+i\,\epsilon)}\qquad\text{for~}\nu_j > 0\,,
\end{align}
to derive the Feynman parametric representation of this integral,
\begin{align}
\label{eq:feynmanparametericrep}
I(\nu_1,...,\nu_N) &= (-1)^{\nu}\,\Gamma(\nu-L\,d/2) \int \left(\prod^{N}_{j=1}\,\frac{\ud x_j\,x_j^{\nu_j-1}}{\Gamma(\nu_j)}\right)
    \delta\left(1-\sum_{j=1}^{N}x_j\right)
\notag\\
&\quad
\frac{\mathcal{U}^{\,\nu-(L+1)\,d/2}}{\mathcal{F}^{\,\nu-L\,d/2}}\quad(\nu_j>0)\,,
\end{align}
with $\nu=\sum_{j=1}^{N}\nu_j$.

We can include inverse propagators (numerators) with $\nu_j<0$ by employing the identity \cite{Smirnov:2012gma,Borowka:2015mxa}
\begin{align}
\label{eq:inverseprops}
    \frac{1}{{(q_j^{2}-m_j^{2}+i\,\epsilon)}^{\nu_j}}
    &= \left[\frac{\partial^{-\nu_j}}{\partial x_j^{-\nu_j}}\,e^{x_j\,(q_j^{2}-m_j^{2}+i\,\epsilon)}\right]_{x_j=0}\qquad\text{for~}\nu_j\leq 0\,.
\end{align}
Let $\mathcal{N}_+$ be the set of all positive $\nu_j$, $\mathcal{N}_-$ the set of all negative $\nu_j$, and $r=\sum_{j\in N_+}\nu_j$. Then, an integral with positive or negative indices can be written as
\begin{align}
\label{eq:feynmanparametersinverseprops}
I(\nu_1,...,\nu_N)
&=(-1)^{r}\,\Gamma(\nu-L\,d/2)
\int
\left(\prod_{j\in\mathcal{N}_+} \frac{\ud x_j\,x^{\nu_j-1}}{\Gamma(\nu_j)}\!\right)
\delta\!\left(\!1-\sum_{j\in\mathcal{N}_+}x_j\!\right)
\notag\\
&\quad
\left[
\left(\prod_{j\in\mathcal{N}_-} \frac{\partial^{|\nu_j|}}{\partial x_j^{|\nu_j|}}\!\right)
\frac{\mathcal{U}^{\,\nu-(L+1)d/2}}{\mathcal{F}^{\,\nu-L\, d/2}}
\right]_{x_j=0\,\forall\, j\in\mathcal{N}_-}
(\nu_j\neq 0).
\end{align}

Our goal is to combine different integrals sharing a common parent topology into one merged parametric representation.
We therefore wish to base our Feynman parametric integral on the resulting $\mathcal{U}$ and $\mathcal{F}$ polynomials for the parent sector. 
For integrals belonging to subtopologies of the parent sector, this can be achieved by taking derivatives with respect to the Feynman parameters corresponding to the pinched lines without setting them to zero,
\begin{align}
\label{eq:pinch}
    \frac{1}{{(q_j^{2}-m_j^{2}+i\,\epsilon)}^{\nu_j}}
    &= - \int_0^\infty \ud x_j \frac{\partial^{-\nu_j+1}}{\partial x_j^{-\nu_j+1}}\,e^{x_j\,(q_j^{2}-m_j^{2}+i\,\epsilon)}\qquad\text{for~}\nu_j\leq 0.
\end{align}
Here, we use the term ``line'' for a propagator with a positive index.
Let $\mathcal{N}=\{1,\ldots,N\}$ be the set of all indices, $\mathcal{N}_T$ the set of positive indices of the parent sector (parent lines),
$\mathcal{N}_t$ the set of positive indices $\nu_j$ of the current sector (integral lines),
$\mathcal{N}_{\Delta t} = \mathcal{N}_T \setminus\mathcal{N}_t$ (set of pinched lines),
$\mathcal{N}_{\setminus T}=\mathcal{N}\setminus\mathcal{N}_T$ be the set of negative indices of the parent sector (parent numerators),
$r=\sum_{j\in\mathcal{N}_t}\nu_j$ the sum of positive indices of the integral,
and $\Delta t = |\mathcal{N}_{\Delta T}|$ the number of pinched lines.
We find
\begin{align}
\label{eq:feynmanparametersgeneral}
I(\nu_1,...,\nu_N)
&=(-1)^{r+\Delta t}\,\Gamma(\nu-L\,d/2)
\int
\left(\prod_{j\in\mathcal{N}_T} \ud x_j\!\right)
\left(\prod_{j\in\mathcal{N}_t} \frac{x^{\nu_j-1}}{\Gamma(\nu_j)}\!\right)
\delta\!\left(\!1-\sum_{j\in\mathcal{N}_T}x_j\!\right)
\notag\\
&\quad
\left[\left(\prod_{j\in\mathcal{N}_{\setminus T}} \frac{\partial^{|\nu_j|}}{\partial x_j^{|\nu_j|}}\!\right)
\left(\prod_{j\in\mathcal{N}_{\Delta t}} \frac{\partial^{|\nu_j|+1}}{\partial x_j^{|\nu_j|+1}}\!\right)
\frac{\mathcal{U}^{\,\nu-(L+1)d/2}}{\mathcal{F}^{\,\nu-L\, d/2}}
\right]_{x_j=0\,\,\forall\,\, j\in\mathcal{N}_{\setminus T}}
(\nu_j\in\mathbb{Z}).
\end{align}
Note that we allow the pinched lines to appear as numerators i.e.\ $\nu_j\leq 0$ for $j \in \mathcal{N}_{\Delta t}$.
The Symanzik polynomials $\mathcal{U}$ and $\mathcal{F}$ are calculated by taking all indices $\mathcal{N}$ into account.
With the prerequisites in place, we can now formulate algorithm~\ref{alg:finite} to construct linear combinations of integrals, which have a convergent Feynman parametric representation for $\epsilon=0$.

\begin{algorithm}
\caption{\label{alg:finite}
Finite Feynman integrals}
\emph{Input:} Dimensionally regularized multiloop integrals with a common parent sector, possibly involving higher powers of propagators, irreducible numerators, or dimension shifts.\\
\emph{Output:} Linear combinations of the input integrals which are finite, i.e.\ they have a convergent Feynman parametric representation for $\epsilon=0$.
\begin{algorithmic}[1]
\State From the $n_s$ input or ``seed'' integrals, form a general linear combination
\begin{equation}
\label{eq:triallincomb}
    I\,=\,\sum^{n_s}_{i=1} a_i I_i\,,
\end{equation}
where $I_i$ are the seed integrals and $a_i$ are the unknown coefficients.
The $a_i$ are assumed to depend on the kinematic invariants and the dimensional regulator $\epsilon$.

\State  Using \eqref{eq:feynmanparametersgeneral}, write the Feynman parametric representation for each seed integral and bring their linear combination over a common denominator such that
\begin{equation}
\label{eq:togetherform}
I = (-1)^{\nu_0}\,\int\left(\prod_{j\in\mathcal{N}_T}\,\ud x_j\!\right) \delta(1-\sum_{j\in\mathcal{N}_T} x_j)\,\,\mathcal{P}\,\,\frac{\mathcal{U}^{\,\nu_0-(L+1)\,(d_0-2\epsilon)/2}}{\mathcal{F}^{\,\nu_0-L\,(d_0-2\epsilon)/2}}\,
\end{equation}
where $\mathcal{N}_T$ is the set of distinct propagators in the parent sector, $\nu_0$ is the effective number of propagators, and $d_0\in\mathbb{Z}$ the effective number of space-time dimensions to be expanded around. The numerator $\mathcal{P}$ is a homogeneous polynomial in the Feynman parameters,
\begin{equation}
\label{eq:numeratorpoly}
    \mathcal{P}\,=\,\sum_{j} c_j\,M_j(x_1,...,x_{\mathcal{N}_T}),
\end{equation}
where the coefficients $c_j$ are polynomials in $a_i$, the kinematic variables, and $\epsilon$, and $M_j(x_1,...,x_{N_p})$ are monomials in Feynman parameters. 
Note that the numerator polynomial $\mathcal{P}$ in general depends on $\epsilon$ and it is crucial to keep this dependence to produce correct results.
It is sufficient, however, to set $\epsilon=0$ in the exponents of the $\mathcal{U}$ and $\mathcal{F}$ polynomials for the convergence analysis in the following two steps.

\State Check the scaling behaviour of the integrand near an integration boundary using the prescription outlined in \cite{Panzer:2013cha,vonManteuffel:2014qoa}.

\State Make sure a convergent integration of \eqref{eq:togetherform} is not prevented by a rapid growth of the integrand near the boundary. This can be achieved by requiring the coefficients of the offending monomials in the numerator to vanish, which provides constraints on the $a_i$. 

\State Repeat 3-4 until all boundaries are checked.
\end{algorithmic}

At the end of this exercise, we are left with
$I\,=\,\sum^{n_\text{fin}}_{i=1} a_i \,\left( \sum_{j=1}^{n_s} b_{ij}\,I_j \right),$
where $n_\text{fin}\geq0$ is the number of finite integrals found, and $ \sum b_{ij}\,I_j$ are the finite combinations.
\end{algorithm}

\clearpage
\newpage
\begin{figure}
\centering
$I_{1,1}:$~ \raisebox{-.5\height}{\includegraphics[width=0.28\linewidth]{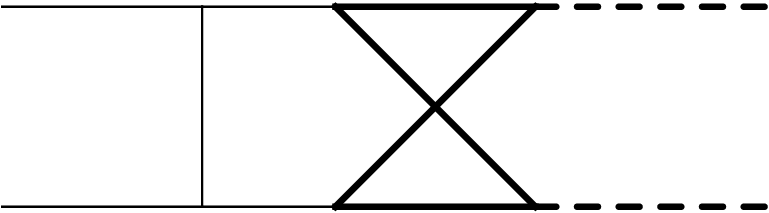}}
\qquad
$I_{2,1}:$~ $(k^2-m_t^2)$\hspace{-1cm}\raisebox{-.5\height}{\includegraphics[width=0.28\linewidth]{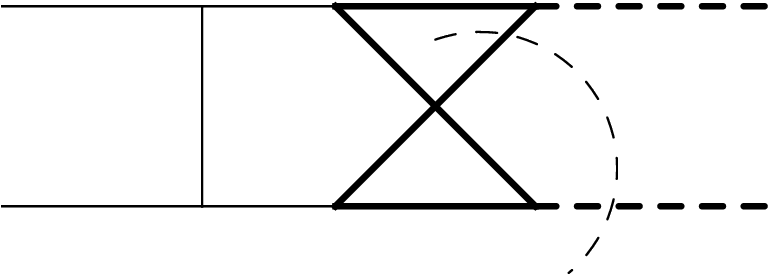}}
\\[0mm]
$I_{3,1}:$~ \raisebox{-.5\height}{\includegraphics[width=0.28\linewidth]{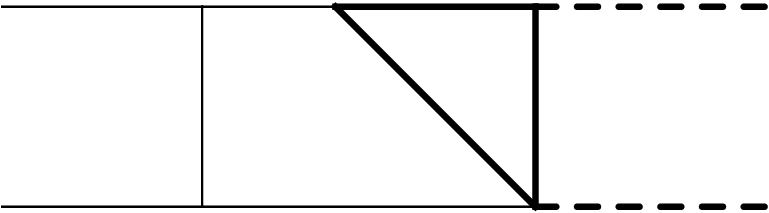}}
\qquad\qquad
$I_{4,1}:$~ \raisebox{-.5\height}{\includegraphics[width=0.28\linewidth]{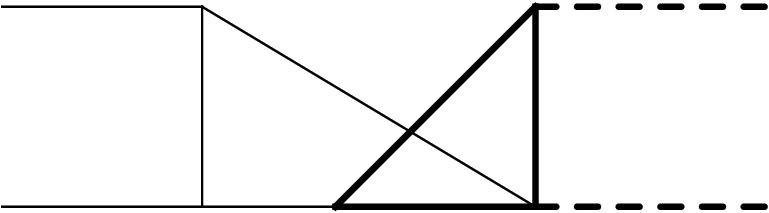}}
\\[2mm]
$I_{5,1}:$~ \raisebox{-.5\height}{\includegraphics[width=0.28\linewidth]{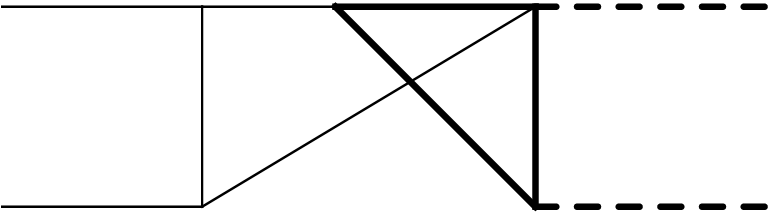}}
\qquad\qquad
$I_{6,1}:$~ \raisebox{-.5\height}{\includegraphics[width=0.28\linewidth]{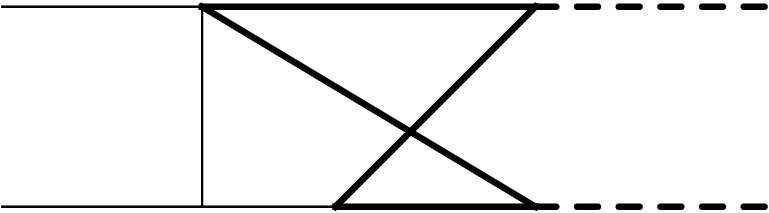}}
\\[2mm]
$I_{7,1}:$~ \raisebox{-.5\height}{\includegraphics[width=0.28\linewidth]{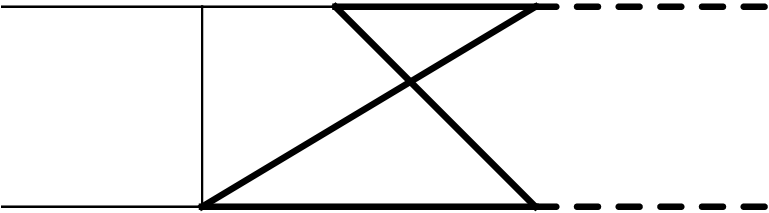}}
\caption{Integrals appearing in \eqref{eq:finitecomb1}.
$I_{1,1}$ is the corner integral of the topology under consideration.
$I_{2,1}$ is a second integral in the topology, but with a numerator $(k^2-m_t^2)$, where $k$ is equal to the difference of the momenta of the edges marked by the thin dashed lines.
Integrals $I_{3,1},I_{4,1},I_{5,1},I_{6,1},I_{7,1}$ belong to subtopologies.
All integrals are defined in $d=4-2\epsilon$ dimensions.}
\label{fig:fincomb1}
\end{figure}
%
\begin{figure}
\centering
$I_{1,2}:$~ $(k^2-m_t^2)$\hspace{-1cm}\raisebox{-.5\height}{\includegraphics[width=0.28\linewidth]{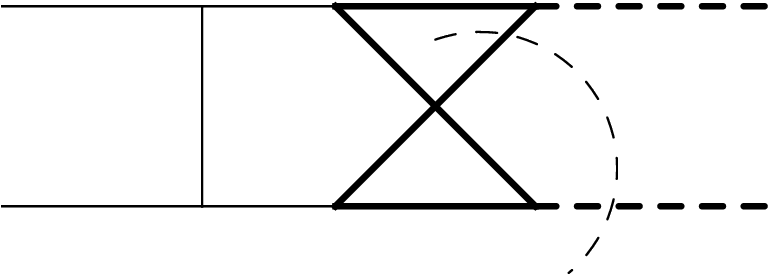}}
\qquad
$I_{2,2}:$~ $(k^2-m_t^2)^2$\hspace{-1cm}\raisebox{-.5\height}{\includegraphics[width=0.28\linewidth]{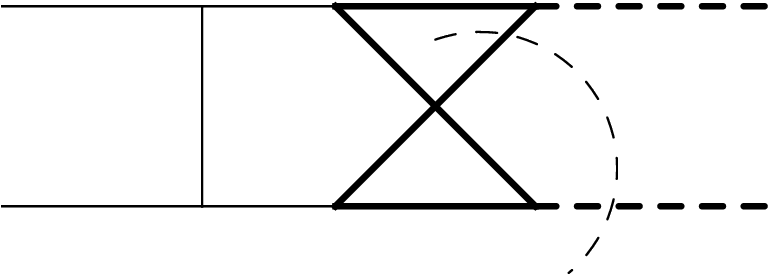}}
\\
$I_{3,2}:$~ $(k^2-m_t^2)$\hspace{-1cm}\raisebox{-.5\height}{\includegraphics[width=0.28\linewidth]{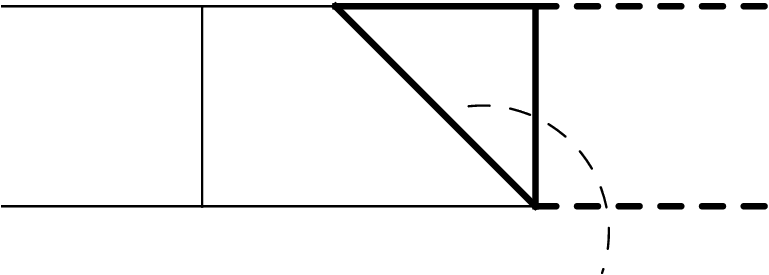}}
\qquad
$I_{4,2}:$~ $(k^2-m_t^2)$\hspace{-1cm}\raisebox{-.5\height}{\includegraphics[width=0.28\linewidth]{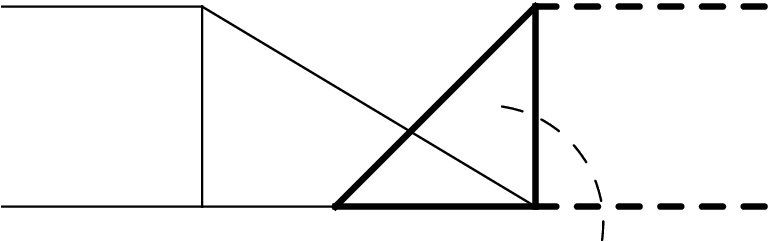}}
\\
$I_{5,2}$:~ $(k^2-m_t^2)$\hspace{-1cm}\raisebox{-.5\height}{\includegraphics[width=0.28\linewidth]{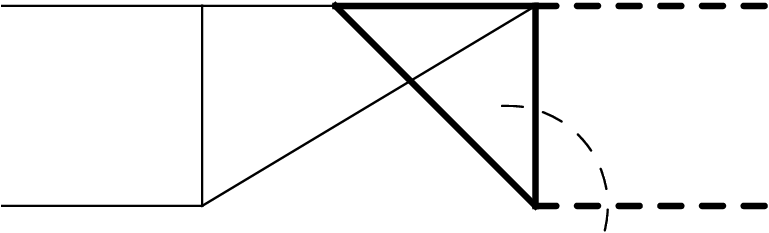}}
\qquad
$I_{6,2}$:~ $(k^2-m_t^2)$\hspace{-1cm}\raisebox{-.5\height}{\includegraphics[width=0.28\linewidth]{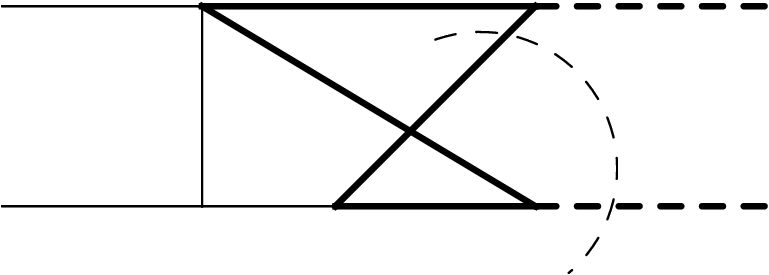}}
\\
$I_{7,2}:$~ $(k^2-m_t^2)$\hspace{-1cm}\raisebox{-.5\height}{\includegraphics[width=0.28\linewidth]{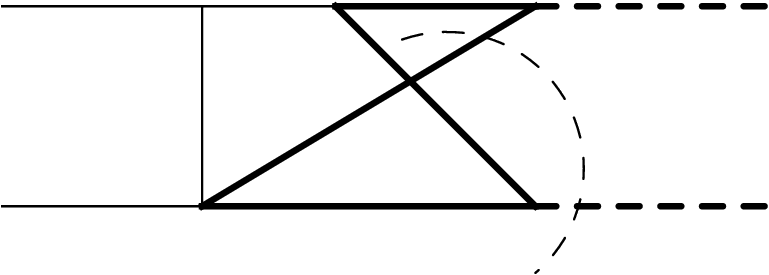}}
\caption{Integrals appearing in \eqref{eq:finitecomb2}.
$I_{1,2}$ is the corner integral of the topology under consideration.
$I_{2,2}$ is a second integral in the topology, but with an extra numerator $(k^2-m_t^2)$ where $k$ is equal to the difference of the momenta of the edges marked by the thin dashed lines.
Integrals $I_{3,2},I_{4,2},I_{5,2},I_{6,2},I_{7,2}$ belong to subtopologies.
All integrals are defined in $d=4-2\epsilon$ dimensions.}
\label{fig:fincomb2}
\end{figure}
As an example, we applied our algorithm to a set of seed integrals including those shown in figure \ref{fig:fincomb1} and obtained the finite linear combination
\begin{equation}
\label{eq:finitecomb1}
    I_{fin,1}\,\,=\,\,s\,(m_z^2-s-t)\,\,I_{1,1}\,+s\,\,I_{2,1}\,\,+s\,\,I_{3,1}\,-s\,\,I_{4,1}\,-s\,\,I_{5,1}\,-(m_z^2-s-t)\,\,I_{6,1}\,-(m_z^2-t)\,\,I_{7,1}\,.
\end{equation}
Allowing for seed integrals with higher numerator rank the algorithm finds, amongst others, the finite linear combination
\begin{equation}
\label{eq:finitecomb2}
    I_{fin,2}\,\,=\,\,s\,(m_z^2-s-t)\,\,I_{1,2}\,+s\,\,I_{2,2}\,\,+s\,\,I_{3,2}\,-s\,\,I_{4,2}\,-s\,\,I_{5,2}\,-(m_z^2-s-t)\,\,I_{6,2}\,-(m_z^2-t)\,\,I_{7,2}\,,
\end{equation}
with the constituent integrals given in figure \ref{fig:fincomb2}.
One can see that $I_{fin,1}$ and $I_{fin,2}$ look very similar. In fact, it is straightforward to see that for any finite linear combination, an additional numerator can be added while keeping the integral IR finite.
Through power counting one can see, that the additional numerator does not introduce a UV divergence in our present example.
However, linear combinations obtained simply by augmenting the existing integrals with additional numerators aren't the only possibilities at higher numerator rank.
Indeed, we observe that generally the number of finite linear combinations increases with the numerator rank.

\subsection{Numerical performance}
One can try to express the amplitude in terms of finite linear combinations, which are defined in $4-2\epsilon$ dimensions and have at most additional numerators.
In practice, we found it useful to consider integrals with ``dots'' and dimension-shifts as well, primarily for the following reasons:
\begin{itemize}
\item It can happen that already the corner integral of a sector has a UV divergence, which can not be cured by a subsector subtraction. Obviously, a numerator insertion is not going to help.
One could try to use a supersector instead, but this can have other disadvantages such as an unnecessary increase of analytic complexity.

\item Choosing integrals with higher numerator ranks leads to extreme proliferation in the number of terms in the numerator polynomial, often leading to rather large {\pysecdec} libraries that are difficult to compile on GPUs. Our efforts to condense the numerators to a more manageable size resulted in the appearance of spurious poles that often worsened numerical stability.

\item In a slightly different approach, integrals with both numerators and dots can be combined to form finite combinations. These integrals, however, have higher powers of the $\mathcal{F}$ polynomial in the denominator. In our experiments, this led to significantly worse numerical performance in the physical region, where contour deformation is required.
\end{itemize}

\begin{table}
\centering
\begin{tabular}{ |c|c|c|c| } 
 \hline
 Integral & Order in $\epsilon$ & Rel. error & Time(s) \\ 
 \hline\hline
 Divergent integral in figure \ref{fig:dimshiftints:1} & $0$ & $\sim 2\cdot 10^{-3}$ & 45\\
 \hline
 Divergent integral in figure \ref{fig:dimshiftints:2} & $0$ & $\sim 4\cdot 10^{-2}$ & 63\\
 \hline
 Finite integral in $d=6-2\epsilon$, in figure \ref{fig:dimshiftints:3} & $1$ & $\sim 8\cdot 10^{-6}$ & 60\\
 \hline
 Finite integral in $d=6-2\epsilon$ with a dot, in figure \ref{fig:dimshiftints:4} & $1$ & $\sim 8\cdot 10^{-4}$ & 55\\
 \hline
 Finite linear combination in \eqref{eq:finitecomb1} & $1$ & $\sim 1\cdot 10^{-4}$ & 18\\
 \hline
 Finite linear combination in \eqref{eq:finitecomb2} & $0$ & $\sim 5\cdot 10^{-4}$ & 150\\
 \hline
\end{tabular}
\caption{Numerical performance of different non-planar integrals for a physical phase-space point. Timings generated with {\pysecdec} \cite{Borowka:2017idc} using the {\tt QMC} algorithm \cite{Li:2015foa,Borowka:2018goh} on an Nvidia Tesla V100S GPU, with $neval=10^7$.}
\label{tab:performance}
\end{table}

A comparison of numerical performance for different divergent and finite integrals for the first few orders in $\epsilon$ expansion is shown in table \ref{tab:performance}. It is clear that the finite integrals perform significantly better. 
The finite integral in figure \ref{fig:dimshiftints:3} has the lowest exponent for $1/\mathcal{F}$, and unsurprisingly shows the best numerical performance. We can also see that the finite linear combination in \eqref{eq:finitecomb1} is on par with the dimension-shifted integrals, which demonstrates its viability. One interesting point to note is that both linear combinations have integrands with $1/\mathcal{F}^3$ compared to $1/\mathcal{F}^2$ for the dimension-shifted finite integral in figure \ref{fig:dimshiftints:4} while having similar performance. 

We observe the best numerical performance for a combination of both approaches: finite linear combinations and dimension-shifted integrals.
In addition, we choose our finite basis of master integrals so that the $d$-dependence of the denominators appearing in the reduction identities factors out, using the code of \cite{Smirnov:2020quc} (see also \cite{Usovitsch:2020jrk}).
In other words, there are no irreducible denominator factors that are polynomials in both kinematics and $d$ for this choice of master integrals. The definitions of the finite master integrals used in our calculation in terms of divergent integrals are provided in an ancillary file.

\section{Renormalisation and checks}
\label{sec:checks}
\subsection{UV renormalisation and IR subtraction}
\label{sec:scheme}
We expand the bare form factors $A_i$ perturbatively according to
\begin{equation}
\label{eq:perturbexpand}
    A_i = \frac{\alpha_{\mathrm{s},0}}{2\pi}\,A_i^{(1)} + \left(\frac{\alpha_{\mathrm{s},0}}{2\pi}\right)^2 A_i^{(2)} + O(\alpha_\mathrm{s}^3)\,,
\end{equation}
where $\alpha_{\mathrm{s},0}$ is the bare QCD coupling.
Since the LO process already starts at one loop, the two-loop process is effectively an NLO correction.

We first perform UV renormalisation of $\alpha_\mathrm{s}$ in the 5-flavour $\overline{\mathrm{MS}}$ scheme, $n_f=5$, with the top-quark contribution to the gluon self energy subtracted at zero momentum \cite{Beenakker:2002nc} using
\begin{equation}
\label{eq:alphasrenorm}
    \alpha_{\mathrm{s},0} = \alpha_\mathrm{s}\,S_\epsilon^{-1}\,Z_{\alpha_\mathrm{s}}\,{\left({\frac{\mu_R^2}{{\mu_0}^2}}\right)}^{\epsilon}\,,
\end{equation}
where $S_\epsilon = (4\pi)^\epsilon e^{-\gamma_E \epsilon}$, $\gamma_E \approx 0.577$ is Euler's constant, $\mu_\mathrm{R}$ is the renormalisation scale, and $\mu_0$ is the 't Hooft scale introduced in the dimensionally regularized bare amplitude.
The renormalisation constant $Z_{\alpha_\mathrm{s}}$ is given by
\begin{equation}
\label{eq:Zalphasdefs}
    Z_{\alpha_\mathrm{s}} = 1+\frac{\alpha_\mathrm{s}}{2\pi}\,\delta Z_{\alpha_\mathrm{s}}\,+\,\mathcal{O}(\alpha_\mathrm{s}^2),
    \quad
    \delta Z_{\alpha_\mathrm{s}} = -\frac{1}{\epsilon}\,\beta_0\,+\,\frac{1}{\epsilon}\,\left(\frac{2}{3}\,T_F\,\left({\frac{\mu_R^2}{m_t^2}}\right)^{\epsilon}\right),
\end{equation}
where 
\begin{equation}
\label{eq:casimirs}
    \beta_0 = \frac{11\,C_A - 4\,T_F\,n_f}{6},\quad C_A = N,\quad C_F = \frac{N^2-1}{2N},\quad T_F = \frac{1}{2}\,.
\end{equation}
We renormalise the top-quark mass in the on-shell scheme. The renormalised top-quark mass is related to the bare mass according to
\begin{equation}
\label{eq:toprenorm}
    m_{t,0}^2\,=\,m_{t}^2\,Z_m,\quad Z_m=1+\frac{\alpha_\mathrm{s}}{2\pi}\,\delta Z_m,\quad
        \delta Z_m\,=\, C_F\,\left(-\frac{3}{\epsilon}\,-4\right)\,\left({\frac{\mu_R^2}{m_t^2}}\right)^{\epsilon}\,.
\end{equation}
In practice, we find it convenient to account for the top-quark mass renormalisation by inserting counterterm vertices in the 1-loop diagrams.
Finally, we take into acount the gluon wave function renormalisation by multiplying the amplitude with $Z_G^{1/2}$ for each external gluon, where the gluon renormalisation constant is defined as
\begin{equation}
\label{eq:gluonrenorm}
    Z_G = 1+\frac{\alpha_\mathrm{s}}{2\pi}\,\left(-\frac{2}{3}\,T_F\,\left({\frac{\mu_R^2}{m_t^2}}\right)^{\epsilon}\right)\,+\,\mathcal{O}(\alpha_\mathrm{s}^2).
\end{equation}
This gives us the renormalised form factors
\begin{equation}
\label{eq:renormff}
    A_i^\text{ren} = \frac{\alpha_{\mathrm{s}}}{2\pi}\,A_i^{(1),\text{ren}} + \left(\frac{\alpha_{\mathrm{s}}}{2\pi}\right)^2 A_i^{(2),\text{ren}} + \mathcal{O}(\alpha_{\mathrm{s}}^3).
\end{equation}

The IR structure of NLO amplitudes was first predicted by Catani in \cite{Catani:1998bh}.
Here, we perform IR subtraction using the ``$q_\mathrm{T}$ scheme'' described in \cite{Catani:2013tia} with
\begin{align}
\label{eq:iop}
    I_{(1)}(\epsilon) &= I_{(1)}^\text{soft}(\epsilon) + I_{(1)}^\text{collinear}(\epsilon),\\
\label{eq:isoft}
    I_{(1)}^\text{soft}(\epsilon) &= -\frac{e^{\epsilon \gamma}}{\Gamma (1-\epsilon)} \, \left(\frac{\mu_R^2}{s}\right)^\epsilon \, \left( \frac{1}{\epsilon^2} + \frac{i\pi}{\epsilon} + \delta_{q_T}^{(0)} \right)\,C_A,\\
\label{eq:icollinear}
    I_{(1)}^\text{collinear}(\epsilon) &= -\left(\frac{\mu_R^2}{s}\right)^\epsilon \, \frac{\beta_0}{\epsilon} ,
\end{align}
where $ \delta_{q_T}^{(0)}=0$.
The finite remainders are then given by
\begin{equation}
\label{eq:finiteff}
    A_{i}^{(2),\text{fin}} = A_i^{(2),\text{ren}} - A_i^{(1),\text{ren}}\,I_{(1)}(\epsilon)\,.
\end{equation}
We present all of our results for $\mu_R^2=s$.

\subsection{Checks}
We perform the following checks to establish the correctness of our results:

\begin{enumerate}[(i)]
\item We verify our 1-loop amplitude against the literature, specifically the form factors provided in \cite{Davies:2020lpf}. This is essential to make sure we match conventions and to facilitate comparisons for the 2-loop result.

\item We explicitly check that the form factors satisfy the identities in \eqref{eq:ffidentities}. We see an exact algebraic identity at the level of reduced amplitude with symbolic kinematics.

\item We also verify, numerically for a phase space point, that the relations in \eqref{eq:ffcrossing} are satisfied.

\item We check all the finite integrals by numerically evaluating them and comparing them against their explicit definitions in terms of divergent integrals for a phase space point.

\item We observe algebraic pole cancellation for the leading poles, see section~\ref{sec:backsub} for details.
We calculate our amplitude for a Euclidean point, using {\reduze} to generate numerical reductions for the Euclidean point. We verify that spurious $1/\epsilon^3$ poles vanish after integration (15 digits), and that the $1/\epsilon^2$ and $1/\epsilon$ poles match Catani's IR formula \cite{Catani:1998bh} (9 digits for the double pole, 7 digits for the single pole) as shown in the first table of appendix~\ref{sec:numchecks}.

\item We verify that the poles match Catani's IR formula \cite{Catani:1998bh} for a point in the physical region as shown in the second table of appendix~\ref{sec:numchecks}. For each phase-space point we compute, we also automatically check that the poles match the IR formula within our numerical uncertainty.

\item We evaluate the amplitude using an alternate finite basis and compare it against our result from the primary basis. This acts as a very strong check of our calculation since it validates the basis change, the definitions of our finite integrals, and the reliability of their numerical evaluation. We find agreement between the two bases within the expected numerical error, typically within a few percent for the form factors. It must be noted that this alternate basis is numerically a lot less stable and unsuitable for large scale evaluation runs.

\item We compare the axial-axial piece of the amplitude evaluated using Kreimer's  anti-commuting $\gamma_5$ scheme \cite{Kreimer:1989ke,Korner:1991sx} with a separate amplitude calculation utilising Larin's $\gamma_5$ scheme~\cite{Larin:1991tj,Larin:1993tq}. For the latter calculation we avoid the appearance of $\gamma_5$ by expressing all axial-currents in terms of Levi-Civita symbols. Metric tensors obtained from contracting two Levi-Civita symbols are treated as $d$-dimensional. Finally, a finite renormalisation is applied for each non-singlet axial current as required to restore the Ward identities.
We emphasize that a verbatim application of the scheme as described in \cite{Larin:1993tq} is motivated (ignoring e.g.\ higher order $\epsilon$ terms in the symmetry restoration constant), because of the finiteness of our one-loop amplitudes.
Performing two independent amplitude calculations utilising different schemes for the treatment of $\gamma_5$ provides a strong check of our amplitude calculation. We find agreement between the two calculations for a physical phase space point within numerical precision.

\item We check that our result reproduces the large top-mass expansion~\cite{Melnikov:2015laa,Caola:2016trd,Campbell:2016ivq} below the top-quark threshold and the small top-mass expansion~\cite{Davies:2020lpf} above; a detailed comparison is presented in the next section.
\end{enumerate}

\section{Results}
\label{sec:results}
Here, we present the results of our calculation and compare them against several approximations available in the literature. In particular, we perform comparisons against the large top-mass expansion~\cite{Melnikov:2015laa,Caola:2016trd,Campbell:2016ivq} as well as the small top-mass power series and Pad\'e expansions~\cite{Davies:2020lpf}.

Let us define the quantities relevant for presentation of our results.
We work in the helicity basis defined by \eqref{eq:helicities}. Concretely, we can write the UV renormalised and IR subtracted helicity amplitudes with incoming helicities $\lambda_1,\lambda_2$ and outgoing helicities $\lambda_3,\lambda_4$ as
\begin{align}
\label{eq:helamps}
\mathcal{M}^{\text{fin}}_{\lambda_1 \lambda_2 \lambda_3 \lambda_4} =  \mathcal{M}^{\text{fin}}_{\mu \nu \rho \sigma} \epsilon_{\lambda_1}^\mu(p_1)
\epsilon_{\lambda_2}^\nu(p_2)
\epsilon_{\lambda_3}^{*\rho}(p_3)
\epsilon_{\lambda_4}^{*\sigma}(p_4).
\end{align}
The amplitudes are expanded as
\begin{align}
\label{eq:helampexpand}
\mathcal{M}^{\text{fin}}_{\lambda_1 \lambda_2 \lambda_3 \lambda_4} = \left( \frac{\alpha_\text{s}}{2\pi} \right) \mathcal{M}_{\lambda_1 \lambda_2 \lambda_3 \lambda_4}^{(1)} + \left( \frac{\alpha_\text{s}}{2\pi} \right)^2 \mathcal{M}_{\lambda_1 \lambda_2 \lambda_3 \lambda_4}^{(2)} + \mathcal{O}(\alpha_\text{s}^3).
\end{align}
To prepare the summation over polarisations, we consider contributions to the squared 1-loop helicity amplitudes $\mathcal{V}^{(1)}$, and to the interference between the 1-loop and 2-loop helicity amplitudes $\mathcal{V}^{(2)}$ defined as
\begin{align}
\label{eq:interf1loop}
    \mathcal{V}_{\lambda_1 \lambda_2 \lambda_3 \lambda_4}^{(1)} &= \mathcal{M}_{\lambda_1 \lambda_2 \lambda_3 \lambda_4}^{*(1)}\mathcal{M}_{\lambda_1 \lambda_2 \lambda_3 \lambda_4}^{(1)}, \\
\label{eq:interf2loop}
    \mathcal{V}^{(2)}_{\lambda_1 \lambda_2 \lambda_3 \lambda_4} &= 2\, \mathrm{Re} \left( \mathcal{M}_{\lambda_1 \lambda_2 \lambda_3 \lambda_4}^{*(1)}\,\mathcal{M}_{\lambda_1 \lambda_2 \lambda_3 \lambda_4}^{(2)} \right).
\end{align}
Note that for our numerical results, we include only the pure top-quark contributions of class A computed here, both in the amplitudes and in the interference terms.
There are 36 helicity amplitudes in total for the $gg \rightarrow ZZ$ process, fulfilling various relations; see section~\ref{sec:form_factors}. In order to further condense the presentation of our results, we average over the helicities of the incoming gluons and define the quantities
\begin{equation}
\label{eq:helampsums}
\mathcal{V}^{(i)}_{\lambda_3 \lambda_4} = \frac{1}{4} \sum_{\lambda_1,\lambda_2} \mathcal{V}^{(i)}_{\lambda_1 \lambda_2 \lambda_3 \lambda_4}\quad \mathrm{and}\quad \mathcal{V}^{(i)} = \sum_{\lambda_3,\lambda_4} \mathcal{V}^{(i)}_{\lambda_3 \lambda_4},\quad(i=1,2),
\end{equation}
where $\lambda_1,\lambda_2\in \{+,-\}$, and $\lambda_3,\lambda_4\in \{+,-,0\}$.
In the results shown, we choose our electroweak couplings as
\begin{align}
\label{eq:ewcouplings}
    G_F &= 1.1663787 \cdot 10^{-5} \ \mathrm{GeV}^{-2}\,,\notag\\
    m_Z &= 91.1876\ \mathrm{GeV}\,,\notag\\
    m_W^2/m_t^2 &= 14/65\,,
\end{align}
where the Fermi constant $G_F$ and Z boson mass $m_Z$ are fixed according to \cite{Zyla:2020zbs}. In our calculation, we fix $m_Z^2/m_t^2 = 5/18$; inserting the value of $m_Z$ from \eqref{eq:ewcouplings} implies that $m_t = 173.016\ \mathrm{GeV}$ and $m_W = 80.296\ \mathrm{GeV}$. The weak mixing angle is fixed according to $\sin(\theta_W) = \sqrt{1 - m_W^2/m_Z^2}$. Note, however, that the only mass ratio fixed in the computationally expensive part of our calculation is that of $m_Z^2/m_t^2$; all other mass values and couplings can straightforwardly be varied in our code. All results are presented at renormalisation scale $\mu^2 = s$.

For numerical evaluation of the integrals appearing in our amplitude we apply sector decomposition and integrate using the quasi-Monte Carlo (QMC) algorithm first applied to sector decomposed Feynman integrals in \cite{Li:2015foa}, as implemented in the program {\pysecdec}~\cite{Borowka:2017idc,Borowka:2018goh}. For a review of QMC methods from a mathematical perspective see, for example, \cite{DickKuoSloan2013}. 
We separately evaluate terms appearing in the form factors of our amplitude according to their colour factor ($C_F$ or $C_A$) and whether they form part of the vector-vector $(v_t^2)$ or axial-axial $(a_t^2)$ contribution. 
For each phase-space point we aim to obtain percent level or better precision for each of the $A_i$ form factors, for each colour structure and for the vector-vector and axial-axial pieces separately.
To present our results, we then rotate to the helicity basis defined in section~\ref{sec:form_factors}. 
In order to improve the efficiency of this approach, the target precision of each integral is set according to its contribution to the uncertainty on the form factors using a variant of the algorithm presented in \cite{Borowka:2016ypz}. 
For most phase-space points, the time required to obtain this precision varies between 90 minutes and 24 hours on 2 Nvidia Tesla V100 GPUs.
This time is completely dominated by the numerical integration of the master integrals;
the time to evaluate the coefficients is basically negligible in this context, see section~\ref{sec:ibp}.
The result of each integral for a given phase-space point is shared between all form factors, colour structures and vector/axial pieces. 
We observe that requiring percent level precision on all of the form factors individually typically results in most of them being obtained to per mille or better precision.
The resulting precision obtained for the interference terms $\mathcal{V}^{(2)}_{\lambda_3 \lambda_4}$ is per mille or better as well.
We expect that a further performance improvement can be achieved by optimising the sampling of the integrals according only to their contribution to the numerical error of the interferences rather than the individual unphysical form factors.
Table~\ref{tab:helamps} shows our numerical results for the independent helicity amplitudes for a physical point in phase space.
The same phase space point is used for the second table in Appendix~\ref{sec:numchecks}, where also the corresponding ($\gamma_5$ scheme dependent) values for the form factors $A_1,\ldots,A_{20}$ are shown.
We would like to emphasize that all of the plots below actually show error bars for our numerical results; however, the errors are too small to be visible in the plots.

\begin{table}
    \centering
    {\small
    \begin{tabular}{|c|c|c|}
        \hline
        $\lambda_1, \lambda_2, \lambda_3, \lambda_4$ & $\mathcal{M}^{(1)}_{\lambda_1 \lambda_2 \lambda_3 \lambda_4}$ (1-loop) & $\mathcal{M}^{(2)}_{\lambda_1 \lambda_2 \lambda_3 \lambda_4}$ (2-loop) \\[2pt]
         \hline\hline
         $++++$ & $0.1337854(1) - 0.0286060(1) \,i$ & $3.15549(8) + 0.47235(8) \,i$ \\
         \hline
         $+++-$ & $0.0015573(1) + 0.0052282(1) \,i$ & $0.15950(7)+0.14052(8) \,i$ \\
         \hline
         $+-+-$ & $-0.01512820(8) - 0.01060416(8) \,i$ & $-0.38609(7) + 0.10539(7) \,i$ \\
         \hline
         $-+++$ & $-0.0291599(1) - 0.0062178(1) \,i$ & $-0.46990(8) + 0.40207(8) \,i$ \\
         \hline
         $+++0$ & $0.0292668(5) + 0.0212966(5) \,i$ & $1.1248(2) - 0.0805(2) \,i$ \\
         \hline
         $+-+0$ & $-0.0643073(5) - 0.0459584(5) \,i$ & $-1.4803(2) + 0.4940(2) \,i$ \\
         \hline
         $++00$ & $0.910006(2) + 1.132536(2) \,i$ & $17.2585(6) + 29.5669(6) \,i$ \\
         \hline
         $+-00$ & $0.355092(2) + 0.404469(2) \,i$ & $10.2869(5) - 1.0571(6) \,i$ \\
         \hline
    \end{tabular}
    }
    \caption{Top-quark contributions to the helicity amplitudes for $gg\to ZZ$ in \eqref{eq:helampexpand}. The results are given for the physical phase space point $s/m_t^2=142/17$, $t/m_t^2=-125/22$, $m_Z^2/m_t^2=5/18$, $m_t=1$ and include only the new contributions of class A defined in section~\ref{sec:diags}. The numbers in parentheses denote the uncertainty in the last digit.}
    \label{tab:helamps}
\end{table}

\begin{figure}[b]
\centering
\includegraphics[height=55mm]{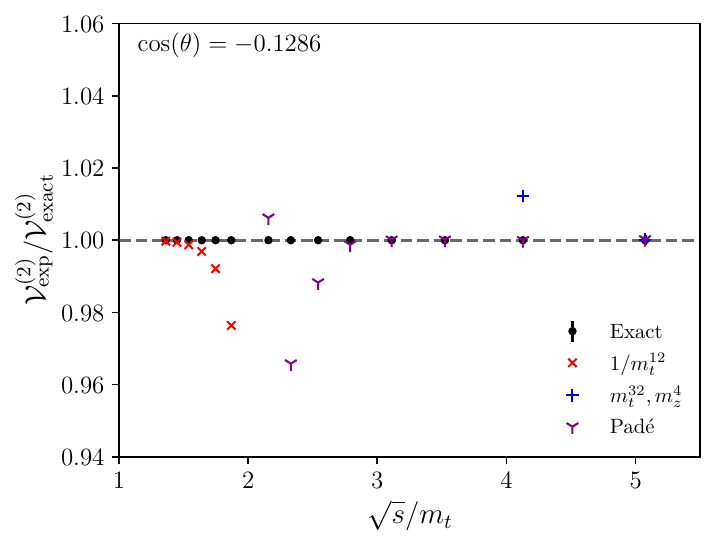}
\caption{Comparison of the $\sqrt{s}$ dependence of the unpolarised interference $\mathcal{V}^{(2)}$ with expansion for large and small top-quark mass~\cite{Davies:2020lpf} at fixed $\cos(\theta)=-0.1286$.}
\label{fig:comarisonbeta}
\end{figure}
In figure \ref{fig:comarisonbeta} we show a comparison of our calculation against the large top-mass and the small top-mass expansions as well as a Pad\'{e} improved small top-mass expansion for a fixed value of $\cos(\theta)$, with $\theta$ being the scattering angle as defined in \eqref{eq:momentumchoice}. 
The plot shows that our calculation agrees very well with the expansions in the relevant regions, which is an important check of our result.
For the smallest value of $\sqrt{s}$, corresponding to $\sqrt{s}=235$ GeV, our result agrees with the large top-mass expansion to better than per mille.
Similarly, for the largest value of $\sqrt{s}$, corresponding to $\sqrt{s}=878$ GeV, both the small top-mass expansion and the Pad\'{e} improved result agree with our result at the sub-per mille level.
Moreover, the best available expansions are in fact capable of reproducing the exact result within a few percent precision for the central scattering angle considered here, except for energies close to the top-quark threshold at $\sqrt{s}=2m_t$.
For the small top mass expansion, we see that the Pad\'e approximation provides a drastic improvement with respect to the power series approach: while the power series data is visible within the plotted range only for the two highest energies and diverges substantially from the exact result for smaller values of $\sqrt{s}$, the Pad\'e approximation agrees very well with our result down to much lower energies.

\begin{figure}
\centering
\includegraphics[height=50mm]{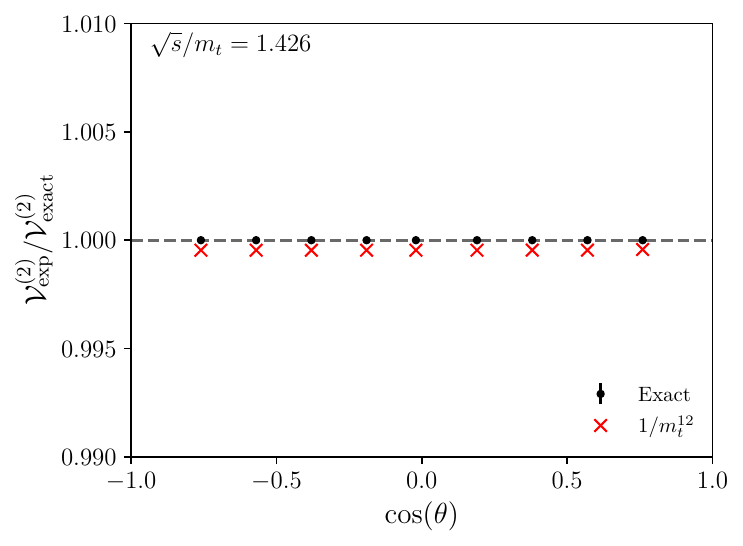}
\includegraphics[height=50mm]{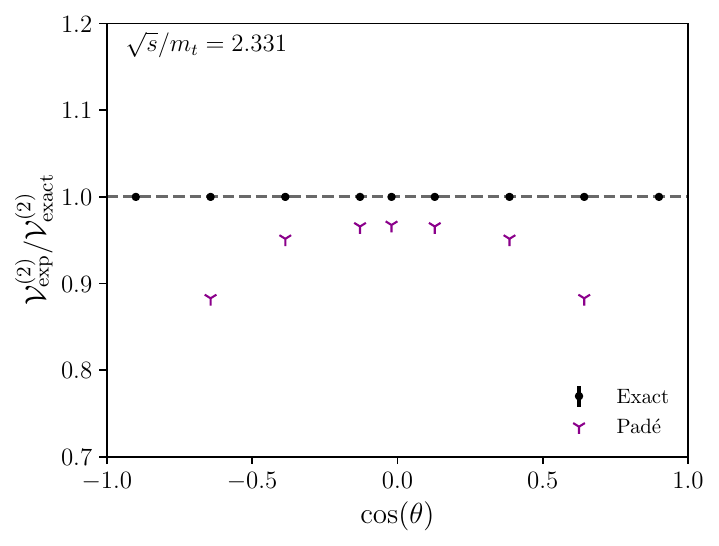}
\includegraphics[height=50mm]{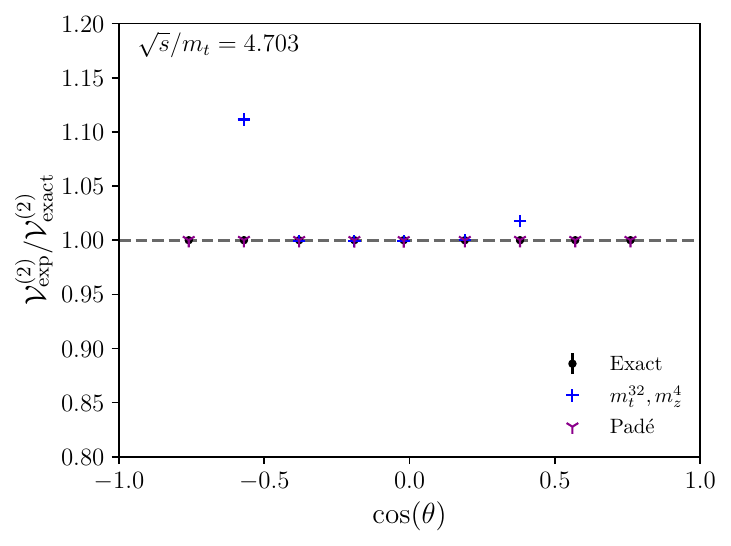}
\caption{Comparison of the $\cos(\theta)$ dependence of the unpolarised interference $\mathcal{V}^{(2)}$ with the results expanded in the limit of large top-quark mass for $\sqrt{s} = 247$ GeV (Top Left Panel) and  small top-quark mass for $\sqrt{s} = 403$ GeV (Top Right Panel) and $\sqrt{s} = 814$ GeV (Bottom Panel).}
\label{fig:comparisontheta}
\end{figure}
%
\begin{figure}
\centering
\includegraphics[height=48mm]{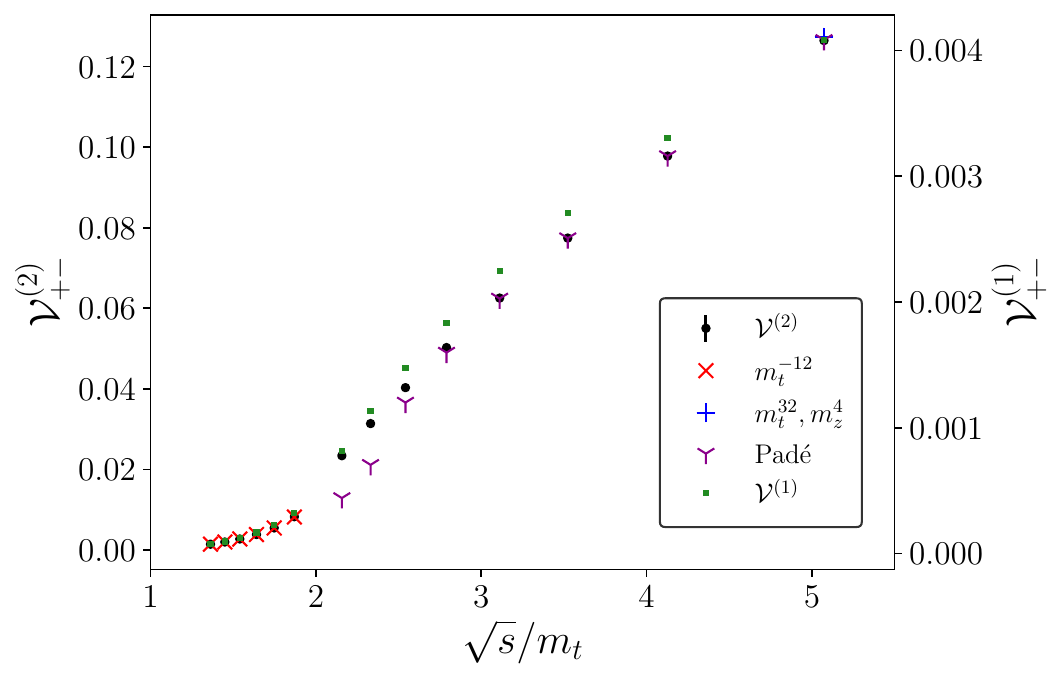}
\includegraphics[height=48mm]{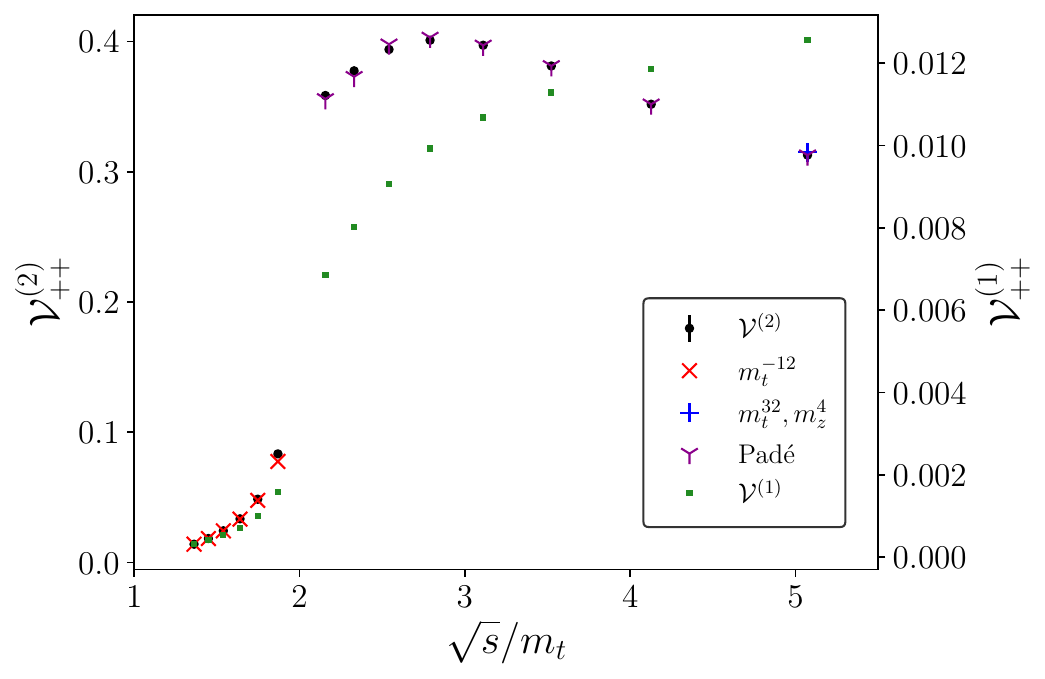}\\
\includegraphics[height=48mm]{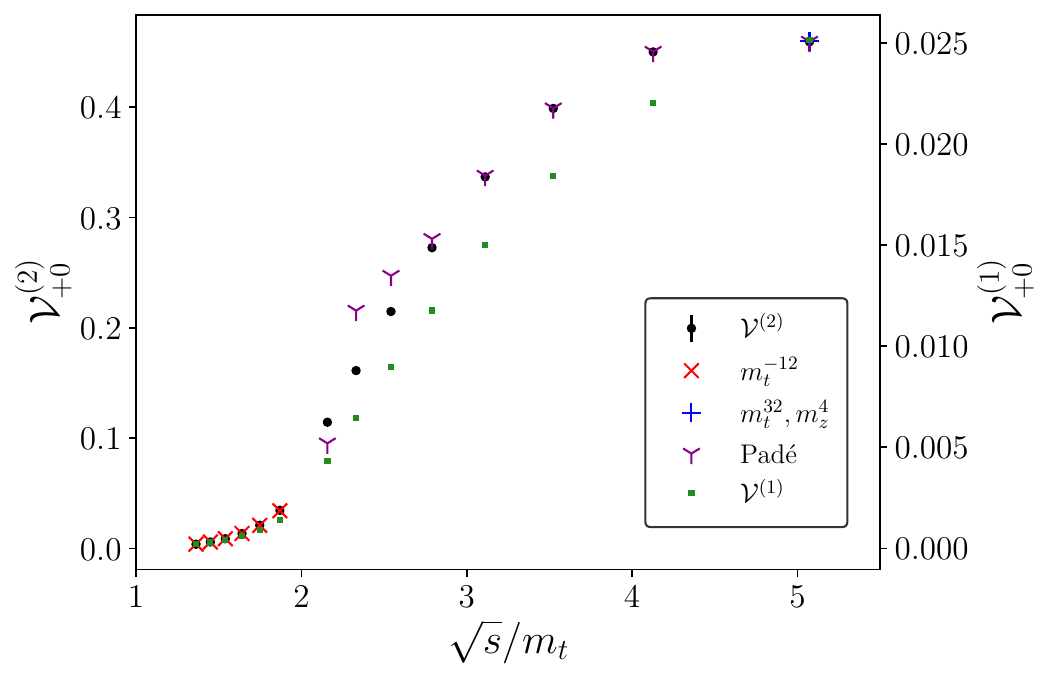}
~~\includegraphics[height=48mm]{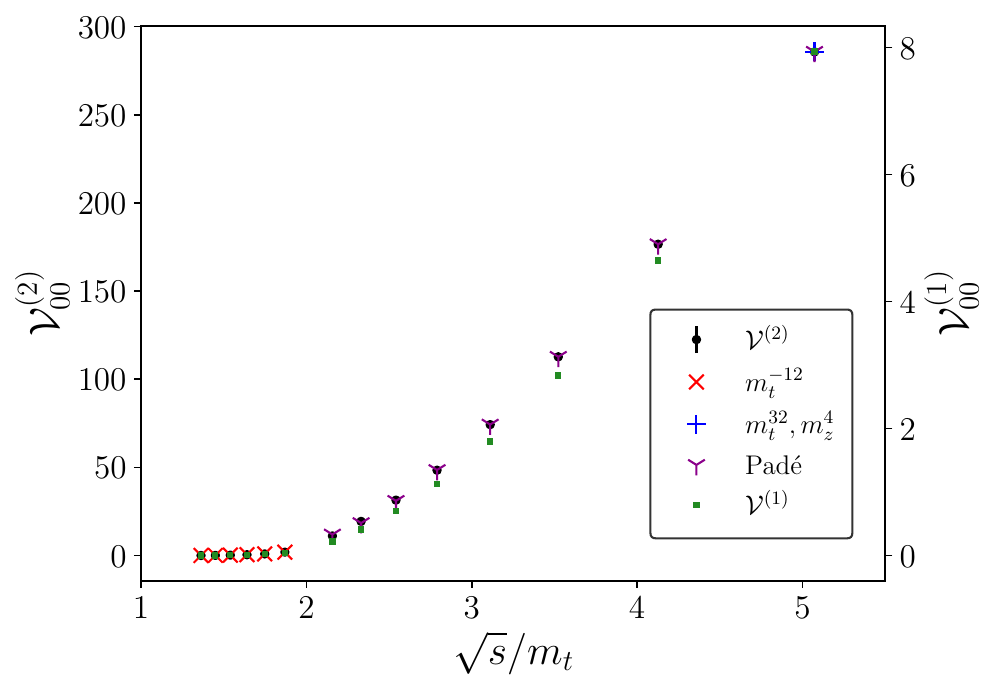}~
\caption{The $\sqrt{s}$ dependence of 1-loop and 2-loop interferences for polarised $ZZ$ production in gluon fusion at $\cos(\theta)=-0.1286$.}
\label{fig:helampsbeta}
\end{figure}

We note that the quantitative level of agreement between the exact results and the approximations depends greatly on the details of the scheme in which the comparison is performed.
For example, to convert the finite 2-loop interference term $\mathcal{V}^{(2)}$ from the ``$q_\mathrm{T}$ scheme'' used in this article (see section~\ref{sec:scheme}) to Catani's original convention \cite{Catani:1998bh} which has also been used in eq.\ 13 of \cite{Davies:2020lpf}, at scale $\mu_R^2 = s$ we must subtract $\pi^2 C_A \mathcal{V}^{(1)}$ (for the real part). The size of the shift resulting from the difference between the two subtraction schemes is comparable to the 2-loop interference terms themselves and can therefore significantly alter the shape of the corrections, their overall size and the level of agreement with the expansions.
Furthermore, the 2-loop curves presented in the following roughly follow the form of the 1-loop curves in the ``$q_\mathrm{T}$ scheme'', while this is generally not the case in Catani's original scheme.
We give explicit examples for these effects in the appendix~\ref{sec:altsubtraction}.

\begin{figure}[b]
\centering
\hspace{-3mm}
\includegraphics[height=46mm,width=.5\linewidth]{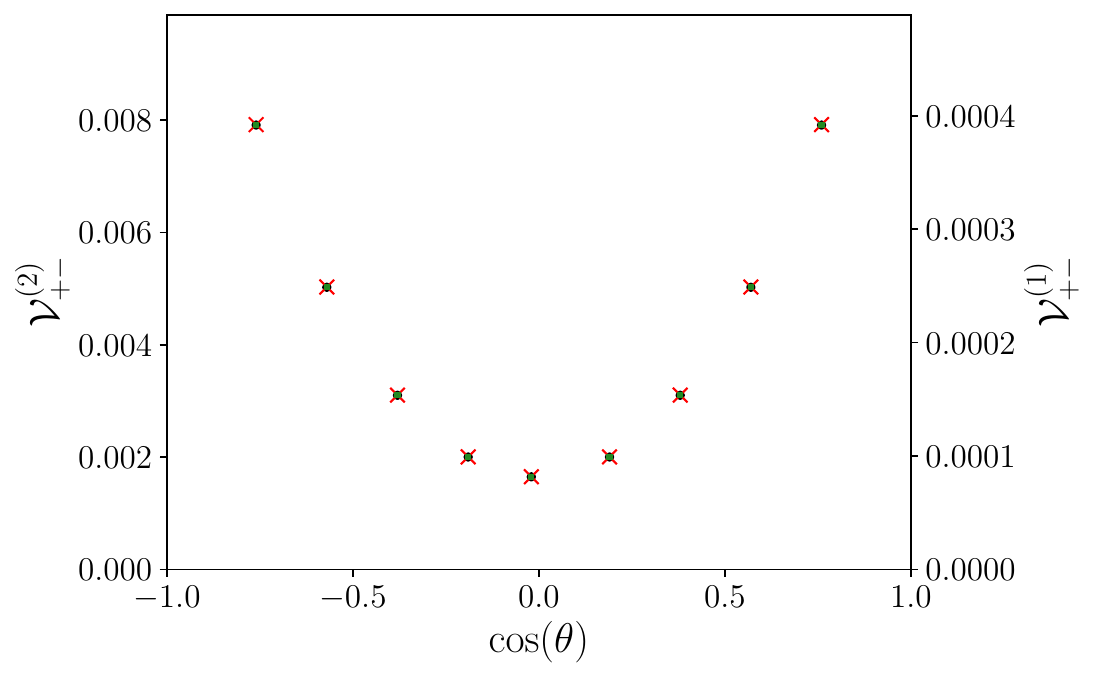}\hspace{0mm}
\includegraphics[height=46mm,width=.5\linewidth]{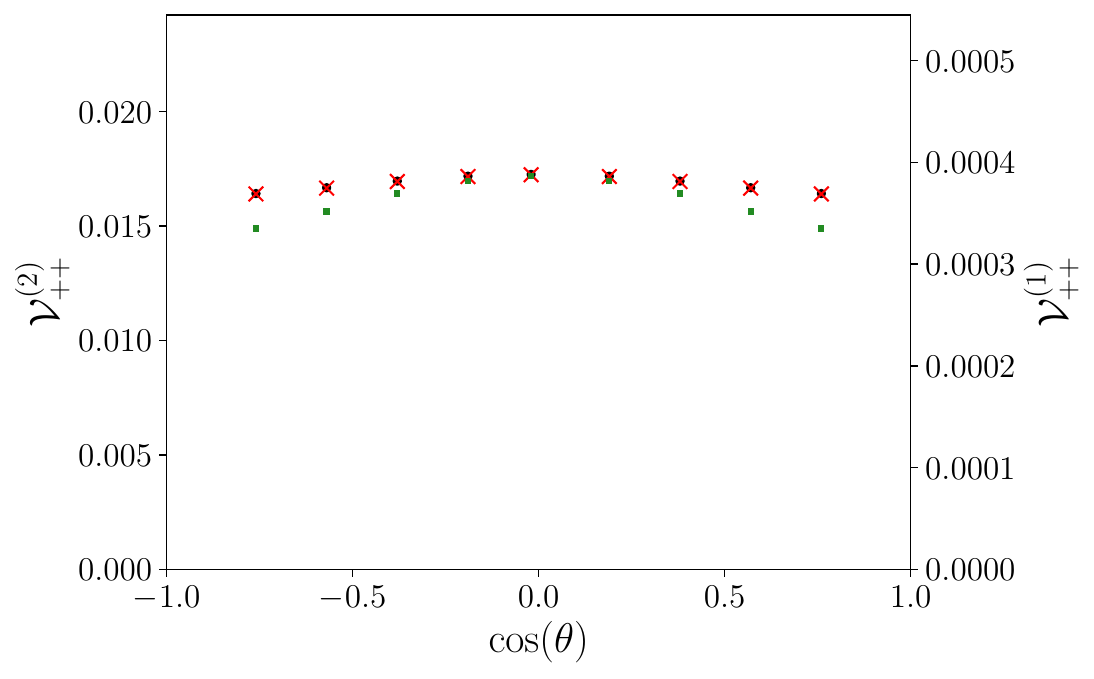}\\[-1.5mm]
\hspace{-3mm}
\includegraphics[height=46mm,width=.5\linewidth]{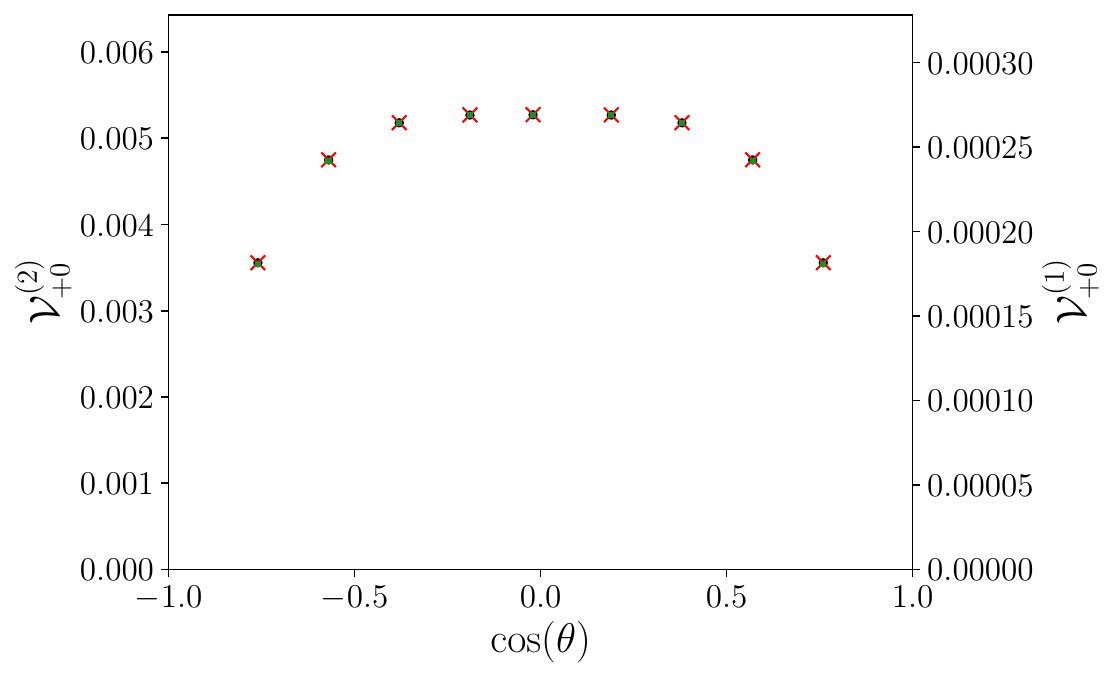}
\includegraphics[height=46mm,width=.5\linewidth]{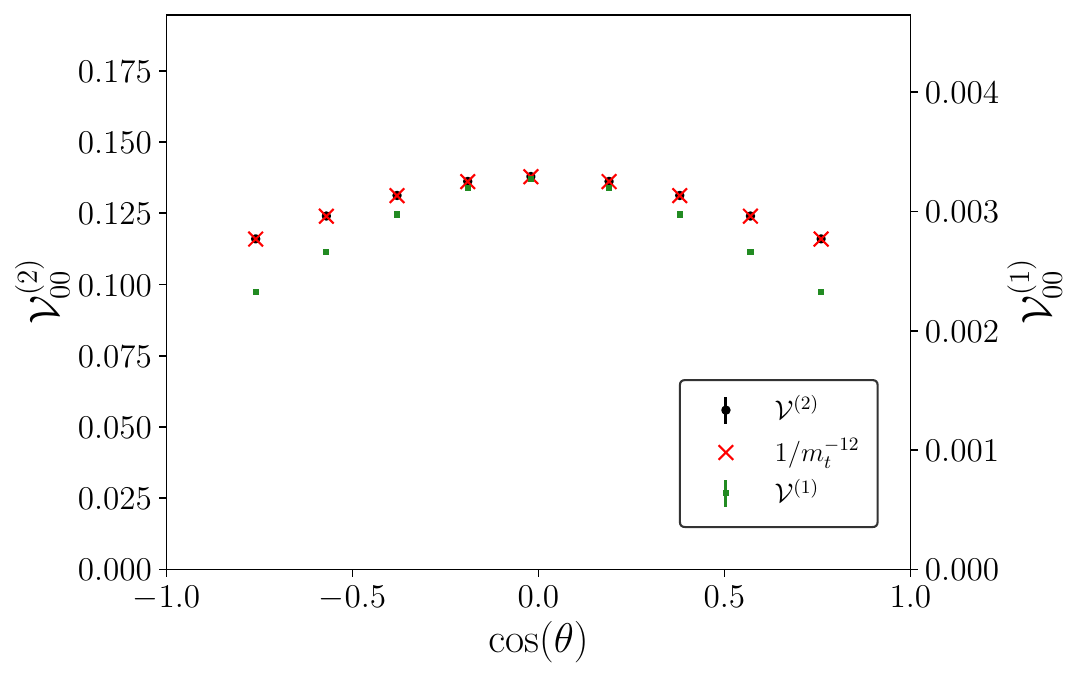}\\[-3mm]
\caption{The $\cos(\theta)$ dependence of 1-loop and 2-loop interferences for polarised $ZZ$ production in gluon fusion at $\sqrt{s}/m_t=1.426$. The large top-quark mass expansion~\cite{Davies:2020lpf} (to order $1/m_t^{12}$) is shown for comparison.}
\label{fig:helampsthetahtl}
\end{figure}
In figure \ref{fig:comparisontheta} we show a comparison of our calculation to the expansions as a function of $\cos(\theta)$ for a fixed value of $\sqrt{s}$. 
It is apparent that the large top-mass expansion is very stable with respect to variation in $\cos(\theta)$ (Top Left Panel). 
The small top-mass power series expansion, on the contrary, diverges rapidly away from $\cos(\theta)=0$.
The breakdown of the small top-mass approximation away from $\cos(\theta)\approx 0$ can be understood from the fact that the expansion is performed in the limit $m_Z^2 \ll m_t^2 \ll s,|t|,|u|$. In particular, for scattering angles $\theta\approx 0$ or $\theta\approx\pi$, the parameters $|t|$ and $|u|$ are not guaranteed to be large compared to $m_t^2$.
The Pad\'e improved expansion substantially cures this problem:
the agreement with our exact result is close to perfect for the high energy samples (Bottom Panel) and good within a few percent for the intermediate energy samples as long as the scattering is relatively central (Top Right Panel).

\begin{figure}
\centering
\hspace{-3mm}
\includegraphics[height=46mm,width=.5\linewidth]{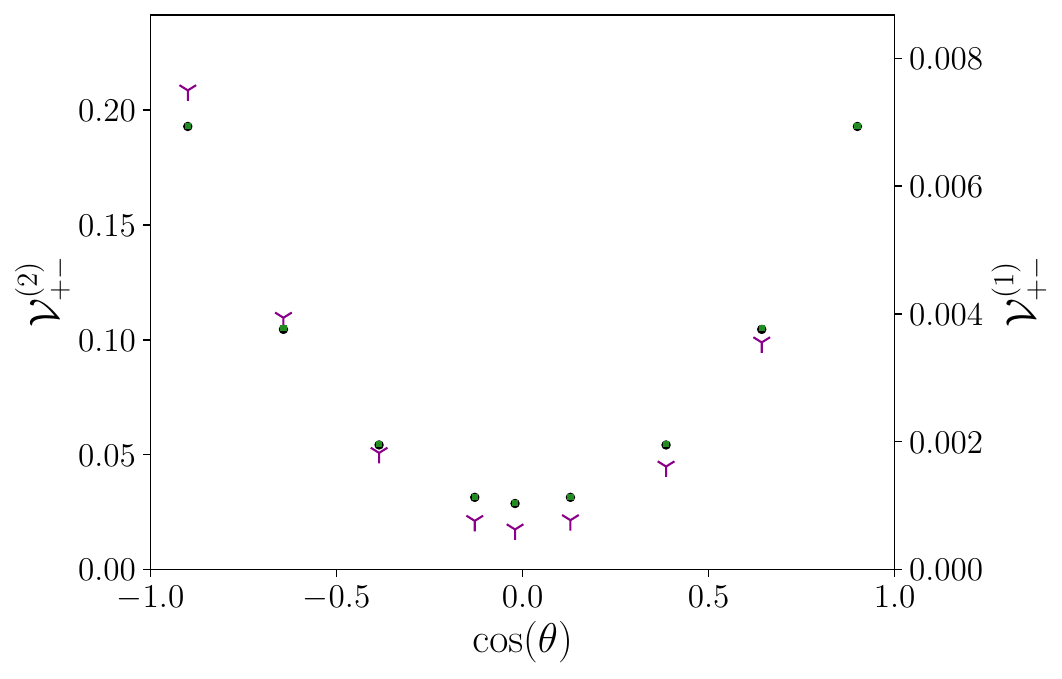}\hspace{0mm}
\includegraphics[height=46mm,width=.5\linewidth]{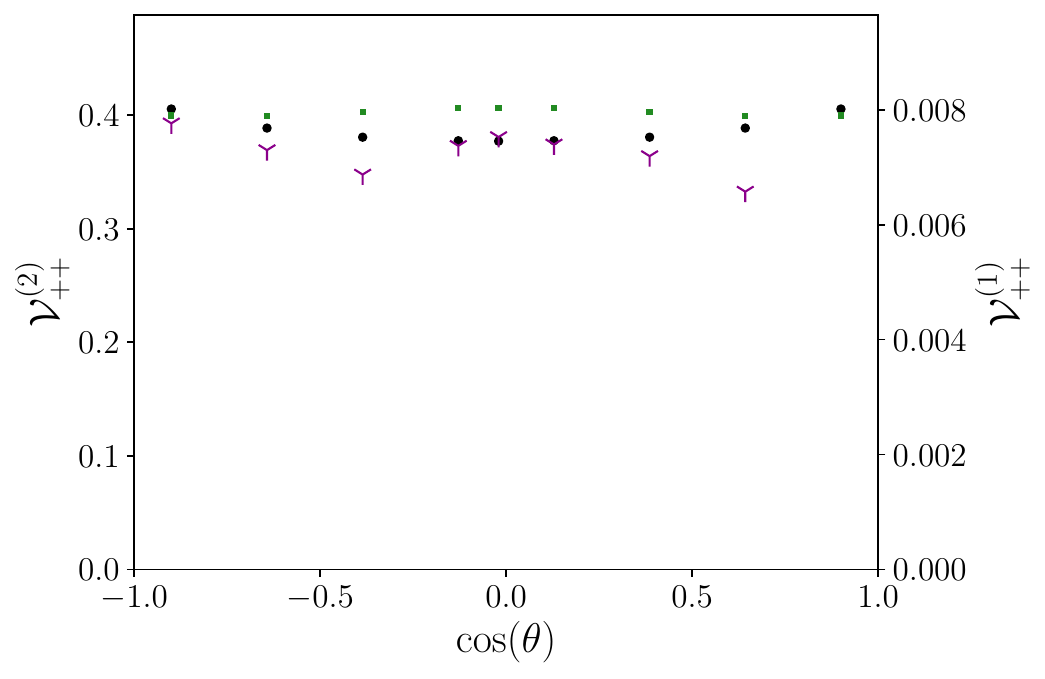}\\[-1.5mm]
\hspace{-3mm}
\includegraphics[height=46mm,width=.5\linewidth]{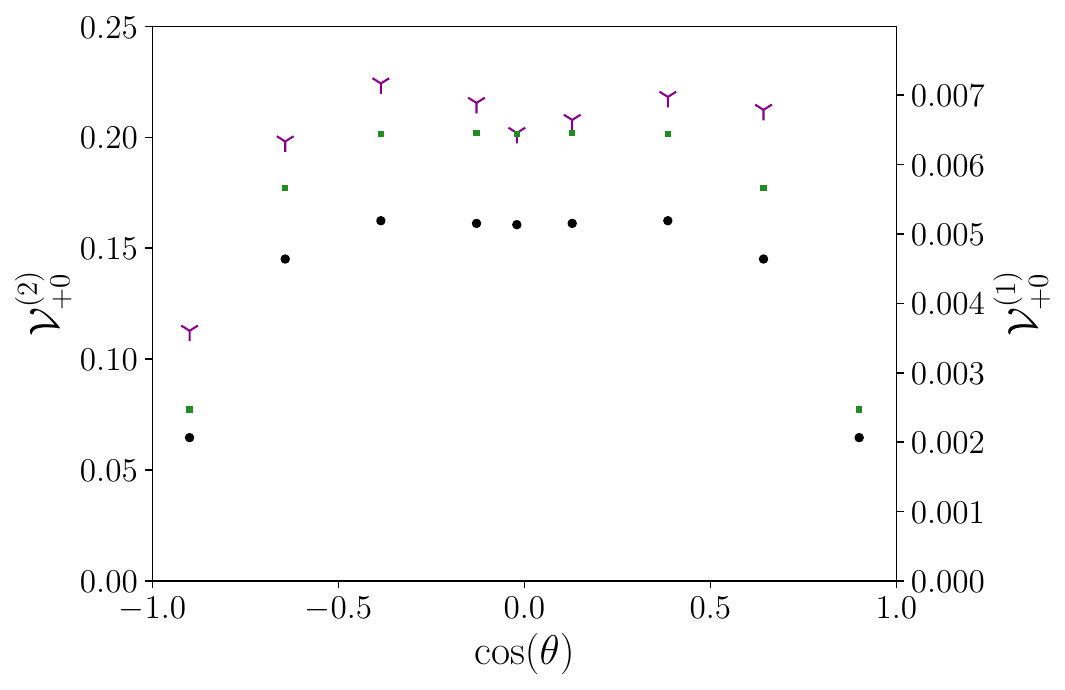}
\includegraphics[height=46mm,width=.5\linewidth]{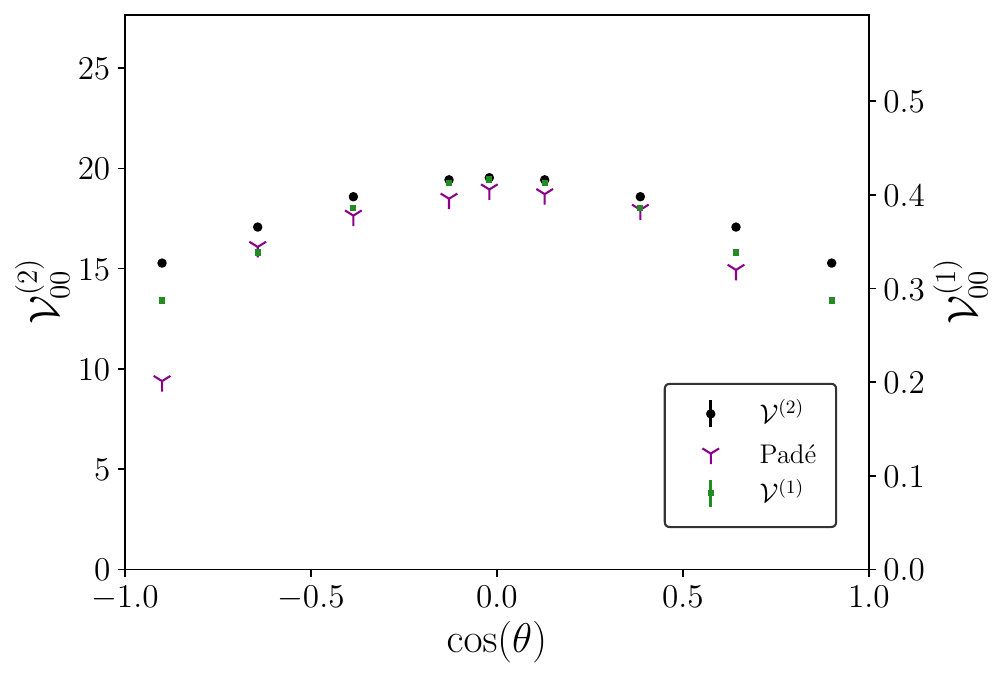}\\[-3mm]
\caption{The $\cos(\theta)$ dependence of 1-loop and 2-loop interferences for polarised $ZZ$ production in gluon fusion at $\sqrt{s}/m_t=2.331$. The Pad\'{e} improved small top-quark mass expansion~\cite{Davies:2020lpf} is shown for comparison.}
\label{fig:helampsthetamid}
\end{figure}
%
\begin{figure}
\centering
\hspace{-3mm}
\includegraphics[height=46mm,width=.5\linewidth]{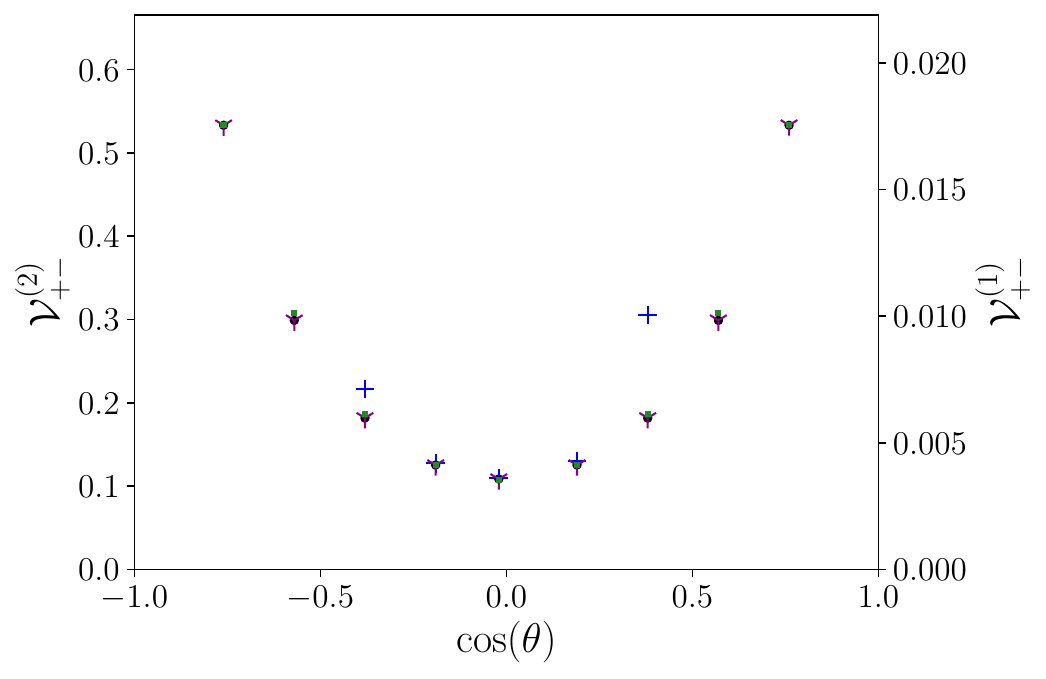}\hspace{-1mm}
\includegraphics[height=46mm,width=.5\linewidth]{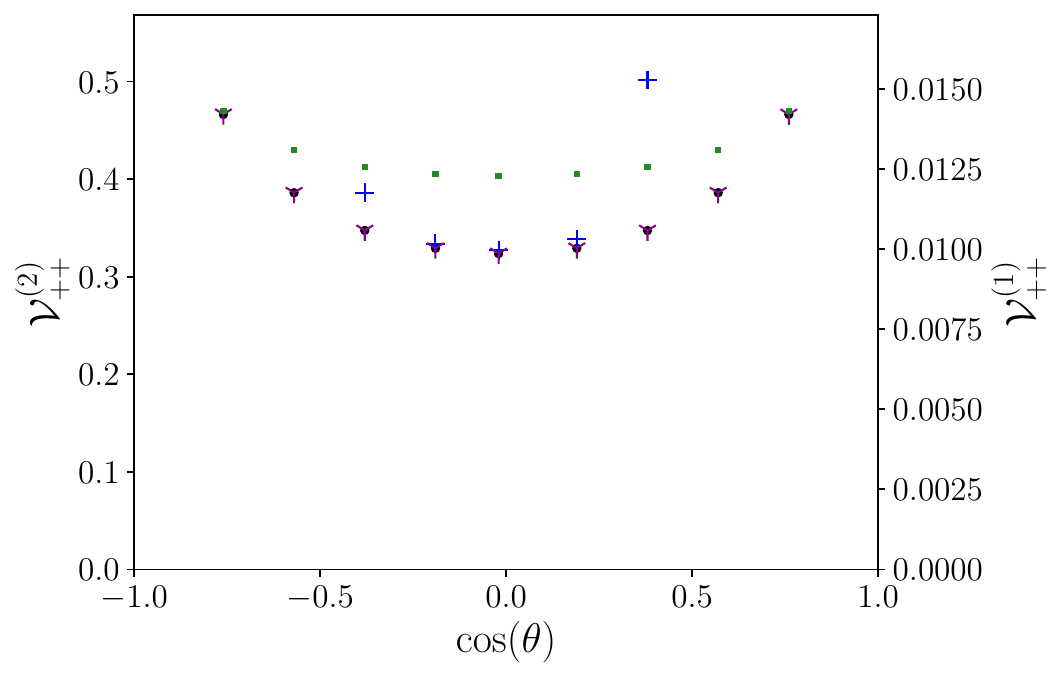}\\[-1.5mm]
\hspace{-3mm}
\includegraphics[height=46mm,width=.5\linewidth]{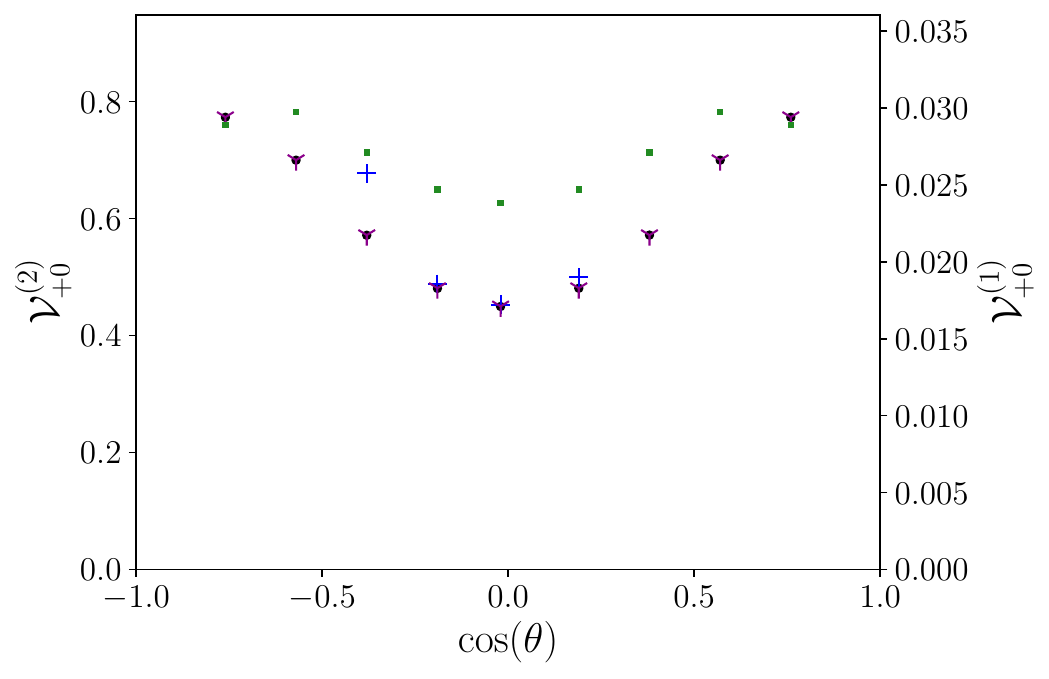}\hspace{-2.5mm}\quad
\includegraphics[height=46mm,width=.49\linewidth]{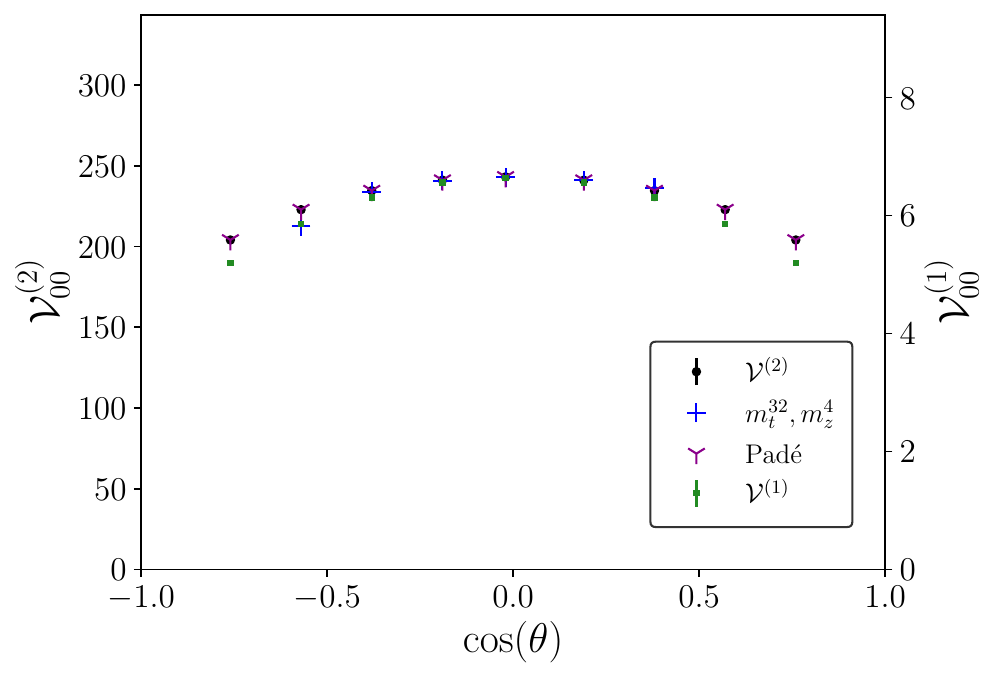}\\[-3mm]
\caption{The $\cos(\theta)$ dependence of 1-loop and 2-loop interferences for polarised $ZZ$ production in gluon fusion at $\sqrt{s}/m_t=4.703$. The small top-quark mass expansion (to order $m_t^{32}$) and Pad\'{e} improved expansion~\cite{Davies:2020lpf} are shown for comparison.}
\label{fig:helampsthetahe}
\end{figure}

Figure \ref{fig:helampsbeta} shows the interferences for specific final state polarizations but averaged over gluon helicities as a function of $\sqrt{s}$ for a fixed value of $\cos(\theta)=-0.1286$. 
We show interference terms at 1-loop, $\mathcal{V}^{(1)}_{\lambda_3 \lambda_4}$, as well as interference terms at two-loops, $\mathcal{V}^{(2)}_{\lambda_3 \lambda_4}$, for different outgoing helicities compared against the expansion results (only at two-loops).
We find good agreement with the expansions in the relevant regions.
We observe that the Pad\'{e} approximation does not agree with the full result equally well for all helicities. Indeed, for the dominant $\mathcal{V}^{(2)}_{00}$ helicity configuration the approximation works well rather close to the top-quark threshold. However, for the suppressed $\mathcal{V}^{(2)}_{+-}$ and $\mathcal{V}^{(2)}_{+0}$ configurations the approximation begins to visibly deteriorate for $\sqrt{s}/m_t \lesssim 3$.
It is interesting to observe that the mode with longitudinal polarisation for both the $Z$ bosons dominates both $\mathcal{V}^{(1)}_{\lambda_3 \lambda_4}$ and $\mathcal{V}^{(2)}_{\lambda_3 \lambda_4}$. 
We also see a rapid increase in $\mathcal{V}^{(1)}_{00}$ and $\mathcal{V}^{(2)}_{00}$ past the $\sqrt{s}=2m_t$ threshold, where the top quarks can be produced on-shell.

In figures \ref{fig:helampsthetahtl}, \ref{fig:helampsthetamid} and \ref{fig:helampsthetahe} we show our results for the polarized interference terms as a function of $\cos(\theta)$ and compare them against expansion results for different fixed values of $\sqrt{s}$.
Note that in many plots the one-loop and two-loop results agree so perfectly when scaled accordingly, that the green points are exactly on top of the black points.
In figure \ref{fig:helampsthetahtl}, we observe that the large top-mass expansion approximates the exact result very well below the $2 m_t$ threshold also as a function $\cos(\theta)$.
For the intermediate energy considered in figure \ref{fig:helampsthetamid}, the Pad\'e result for small top-quark mass agrees overall rather well with the exact results.
As expected from the previous discussion, some deviations are visible for non-dominant final state helicities or non-central scattering.
Further, we note an asymmetry in the $\cos(\theta)$ dependence, which is a consequence of the asymmetric expansion used to construct the Pad\'{e} approximation. In figure \ref{fig:comparisontheta}, this asymmetry is absent by construction since in this unpolarised case the Pad\'{e} approximation is calculated for a fixed value of $\sqrt{s}$ and $p_T$ and used for both the forward and backward directions.
For the high energy in figure \ref{fig:helampsthetahe}, we see excellent agreement between the angular dependence of the Pad\'e result and that of the exact result.
As visible from the figure, this is a significant improvement over the angular dependence of the power series approach to the small mass expansion for less central scattering angles.
\FloatBarrier

\section{Conclusions}
\label{sec:conclusions}
In this paper, we have presented a calculation of the two-loop top-quark corrections for the process $gg\to ZZ$.
Maintaining exact dependence on the top-quark mass, we calculated the helicity amplitudes in terms of finite integrals, which we evaluated using numerical quadrature.
For reduction to master integrals, we employed finite field techniques and syzygies which avoid the introduction of squared propagators (``dots'').
We presented a new computational method to find these syzygies with linear algebra.

We considered finite linear combinations of dimensionally regularized Feynman integrals and presented a novel algorithm to systematically construct them.
These linear combinations possess convergent parametric integral representations for $\epsilon=0$ and are formed from building blocks which may involve irreducible numerators, higher powers of propagators, dimensionally shifted integrals, and subsector integrals.
The resulting parametric integrand can be expanded in $\epsilon$ allowing for direct numerical integration.
We employed {\pysecdec} and found that such finite linear combinations can significantly improve the numerical performance of the quadrature, similar to what was observed in the case of dimensionally shifted integrals with additional dots.
The new approach allows us to stay in a range of integrals which may be considered more natural in the context of the amplitudes themselves.
We emphasize that our method is fully automated and works for arbitrary loop order and number of external legs.

Our results for the two-loop amplitudes show good agreement with the large $m_t$ expansion and the small $m_t$ expansions in the regions, where they are expected to be valid.
At moderate energies and for non-central scattering at higher energies, we find that the small $m_t$ power series expansion differs substantially from our result.
In contrast, the Pad\'e improved results~\cite{Davies:2020lpf} give a very good approximation to our results for a much larger region of phase space.

We observed that the quantitative and even qualitative behavior of the 2-loop finite remainders is rather sensitive to the choice of infrared subtraction terms.
In particular, admixtures of 1-loop contributions may actually dominate the overall behavior of the 2-loop remainders and smooth 2-loop threshold effects.
As a consequence, the level of agreement between the available approximations and our exact result depends significantly on the choice of the subtraction scheme.

Our amplitudes provide the last major building block required to include the full top-quark mass effects in the next-to-leading order cross section for $ZZ$ production in gluon fusion.

\acknowledgments
We are very grateful to Stephan Jahn for providing checks of our calculations and for helping interface our finite integral finder with \texttt{pySecDec}.
We would like to gratefully acknowledge Joshua Davies, Go Mishima, and Matthias Steinhauser for providing detailed expansion results, which enabled the numerical comparisons presented in this article.
We thank Christian Br\o{}nnum-Hansen and Chen-Yu Wang for valuable communications regarding their work~\cite{Bronnum-Hansen:2021olh}, which helped us to fix issues in the numerical tables in the appendix.
We wish to thank Robert M.\ Schabinger for many illuminating discussions concerning the construction of finite integrals
and Yang Zhang for interesting discussions regarding the construction of syzygies.
We acknowledge helpful discussions with Gudrun Heinrich, Nikolas Kauer, Matthias Kerner, Kirtimaan Mohan, Erik Panzer, and C.-P.\ Yuan.
BA thanks the Max Planck Institute for Physics, Munich, for support and hospitality.
BA and AvM were supported in part by the National Science Foundation under Grants No.\ 1719863
and 2013859.
SJ was supported in part by a Royal Society University Research Fellowship under Grant URF/R1/201268.
We gratefully acknowledge support and resources
provided by the Max Planck Computing and Data Facility (MPCDF) and the High Performance Computing Center (HPCC) at Michigan State University.
Our Feynman diagrams were generated using {\tt JaxoDraw} \cite{Binosi:2003yf}, based on {\tt AxoDraw} \cite{Vermaseren:1994je}.
\newpage
\appendix
\section{Numerical checks}
\label{sec:numchecks}
In this appendix, we present details for the numerical pole checks for our basis of finite integrals in Kreimer's anti-commuting $\gamma_5$ scheme.
{For the UV renormalised two-loop form factors prior to IR subtraction, we observe analytical pole cancellation through to order $1/\epsilon^4$ and very precise numerical cancellations at order $1/\epsilon^3$. For the $1/\epsilon^2$ and $1/\epsilon$ poles, the following table shows our results compared against predicted IR poles (\ref{eq:isoft}, \ref{eq:icollinear}) as well as the $\epsilon^0$ term (before IR subtraction) for the Euclidean point $s/m_t^2=-191$, $t/m_t^2=-337$, $m_Z^2/m_t^2=-853$, $m_t=1$. The digits in parentheses for the $\epsilon^0$ term denote the uncertainty in the last digit.}
\begin{center}
{\tiny
\begin{tabular}{ |c|c|c|c| } 
 \hline
 FF & $1/\epsilon^2$ & $1/\epsilon$ & $\epsilon^0$\\
 \hline\hline
 $A_1$ & $+2.436734851\cdot 10^{-1}$ & $+8.212518984 \cdot 10^{-1} +1.531045661\,i$ & $-2.806661(2)+4.18190980(3)\,i$\\
       Pred. & $+2.436734852\cdot 10^{-1}$ & $+8.212518977 \cdot 10^{-1}+1.531045662\,i$ & \\
 \hline
 $A_2$ & $-1.760872097\cdot 10^{-1}$ & $-6.021429768\cdot 10^{-1}-1.106388569\,i$ & $+2.509969(1)-3.07651654(4)\,i$\\
       Pred. & $-1.760872097\cdot 10^{-1}$ & $-6.021429781\cdot 10^{-1}-1.106388569\,i$&\\
 \hline
 $A_3$ & $-3.815946068\cdot 10^{-2}$ & $-7.236587884\cdot 10^{-2}-2.397629627\cdot 10^{-1}\,i$ & $+1.2102(3)\cdot 10^{-2}-3.015063(4)\cdot 10^{-1}\,i$\\
       Pred. & $-3.815946069\cdot 10^{-2}$ & $-7.236587838\cdot 10^{-2}-2.397629627\cdot 10^{-1}\,i$&\\
 \hline
 $A_4$ & $-1.565000574\cdot 10^{-4}$ & $-5.374251500\cdot 10^{-4}-9.833188615\cdot 10^{-4}\,i$ & $+2.18538(3)\cdot 10^{-3}-2.748510(3)\cdot 10^{-3}\,i$\\
       Pred. & $-1.565000575\cdot 10^{-4}$ & $-5.374251489\cdot 10^{-4}-9.833188622\cdot 10^{-4}\,i$&\\
 \hline
 $A_5$ & $+7.608919171\cdot 10^{-4}$ & $+1.926944077\cdot 10^{-3}+4.780824914\cdot 10^{-3}\,i$ & $-1.051486(4)\cdot 10^{-2}+9.052930(4)\cdot 10^{-3}\,i$\\
       Pred. & $+7.608919168\cdot 10^{-4}$ & $+1.926944068\cdot 10^{-3}+4.780824912\cdot 10^{-3}\,i$&\\
 \hline
 $A_6$ & $+7.576619247\cdot 10^{-4}$ & $+2.735071357\cdot 10^{-3}+4.760530273\cdot 10^{-3}\,i$ & $-7.02484(5)\cdot 10^{-3}+1.41435102(3)\cdot 10^{-2}\,i$\\
       Pred. & $+7.576619247\cdot 10^{-4}$ & $+2.735071351\cdot 10^{-3}+4.760530273\cdot 10^{-3}\,i$&\\
 \hline
 $A_7$ & $-1.565000574\cdot 10^{-4}$ & $-5.374251500\cdot 10^{-4}-9.833188615\cdot 10^{-4}\,i$ & $+2.18538(3)\cdot 10^{-3}-2.748510(3)\cdot 10^{-3}\,i$\\
       Pred. & $-1.565000575\cdot 10^{-4}$ & $-5.374251489\cdot 10^{-4}-9.833188622\cdot 10^{-4}\,i$&\\
 \hline
 $A_8$ & $-3.055600405\cdot 10^{-4}$ & $-1.158849558\cdot 10^{-3}-1.919890357\cdot 10^{-3}\,i$ & $+4.35036(1)\cdot 10^{-3}-6.0546699(5)\cdot 10^{-3}\,i$\\
       Pred. & $-3.055600405\cdot 10^{-4}$ & $-1.158849559\cdot 10^{-3}-1.919890357\cdot 10^{-3}\,i$&\\
 \hline
 $A_9$ & $+2.001982671\cdot 10^{-4}$ & $+7.482078266\cdot 10^{-4}+1.257882810\cdot 10^{-3}\,i$ & $-3.07299(1)\cdot 10^{-3}+3.897481(1)\cdot 10^{-3}\,i$\\
       Pred. & $+2.001982671\cdot 10^{-4}$ & $+7.482078292\cdot 10^{-4}+1.257882810\cdot 10^{-3}\,i$&\\
 \hline
 $A_{10}$ & $+3.636573767\cdot 10^{-4}$ & $+1.390161598\cdot 10^{-3}+2.284926686\cdot 10^{-3}\,i$ & $-4.77622(2)\cdot 10^{-3}+7.274828(2)\cdot 10^{-3}\,i$\\
          Pred. & $+3.636573768\cdot 10^{-4}$ & $+1.390161596\cdot 10^{-3}+2.284926686\cdot 10^{-3}\,i$&\\
 \hline
 $A_{11}$ & $+5.388240322\cdot 10^{-6}$ & $-1.272166624\cdot 10^{-4}+3.385531242\cdot 10^{-5}\,i$ & $+1.04254(1)\cdot 10^{-3}-8.20955(1)\cdot 10^{-4}\,i$\\
          Pred. & $+5.388240348\cdot 10^{-6}$ & $-1.272166651\cdot 10^{-4}+3.385531259\cdot 10^{-5}\,i$&\\
 \hline
 $A_{12}$ & $-5.388240322\cdot 10^{-6}$ & $+1.272166624\cdot 10^{-4}-3.385531242\cdot 10^{-5}\,i$ & $-1.04254(1)\cdot 10^{-3}+8.20955(1)\cdot 10^{-4}\,i$\\
          Pred. & $-5.388240348\cdot 10^{-6}$ & $+1.272166651\cdot 10^{-4}-3.385531259\cdot 10^{-5}\,i$&\\
 \hline
 $A_{13}$ & $-3.636573767\cdot 10^{-4}$ & $-1.390161598\cdot 10^{-3}-2.284926686\cdot 10^{-3}\,i$ & $+4.77622(2)\cdot 10^{-3}-7.274828(2)\cdot 10^{-3}\,i$\\
          Pred. & $-3.636573768\cdot 10^{-4}$ & $-1.390161596\cdot 10^{-3}-2.284926686\cdot 10^{-3}\,i$&\\
 \hline
 $A_{14}$ & $-2.001982671\cdot 10^{-4}$ & $-7.482078266\cdot 10^{-4}-1.257882810\cdot 10^{-3}\,i$ & $+3.07299(1)\cdot 10^{-3}-3.897481(1)\cdot 10^{-3}\,i$\\
          Pred. & $-2.001982671\cdot 10^{-4}$ & $-7.482078292\cdot 10^{-4}-1.257882810\cdot 10^{-3}\,i$&\\
 \hline
 $A_{15}$ & $+3.055600405\cdot 10^{-4}$ & $+1.158849558\cdot 10^{-3}+1.919890357\cdot 10^{-3}\,i$ & $-4.35036(1)\cdot 10^{-3}+6.0546699(5)\cdot 10^{-3}\,i$\\
          Pred. & $+3.055600405\cdot 10^{-4}$ & $+1.158849559\cdot 10^{-3}+1.919890357\cdot 10^{-3}\,i$&\\
 \hline
 $A_{16}$ & $+1.898361362\cdot 10^{-4}$ & $+6.165488820\cdot 10^{-4}+1.192775622\cdot 10^{-3}\,i$ & $-2.233448(2)\cdot 10^{-3}+3.11183978(6)\cdot 10^{-3}\,i$\\
          Pred. & $+1.898361362\cdot 10^{-4}$ & $+6.165488809\cdot 10^{-4}+1.192775622\cdot 10^{-3}\,i$&\\
 \hline
 $A_{17}$ & $-4.235989659\cdot 10^{-8}$ & $-1.659620988\cdot 10^{-7}-2.661550798\cdot 10^{-7}\,i$ & $+8.1249(2)\cdot 10^{-7}-8.72727(4)\cdot 10^{-7}\,i$\\
          Pred. & $-4.235989677\cdot 10^{-8}$ & $-1.659621000\cdot 10^{-7}-2.661550810\cdot 10^{-7}\,i$&\\
 \hline
 $A_{18}$ & $-9.857950093\cdot 10^{-8}$ & $-9.594603102\cdot 10^{-7}-6.193932718\cdot 10^{-7}\,i$ & $+4.4198(6)\cdot 10^{-7}-5.632743(5)\cdot 10^{-6}\,i$\\
          Pred. & $-9.857950139\cdot 10^{-8}$ & $-9.594603103\cdot 10^{-7}-6.193932747\cdot 10^{-7}\,i$&\\
 \hline
 $A_{19}$ & $+8.932087549\cdot 10^{-7}$ & $+3.205282901\cdot 10^{-6}+5.612196125\cdot 10^{-6}\,i$ & $-7.43447(5)\cdot 10^{-6}+1.6553816(4)\cdot 10^{-5}\,i$\\
          Pred. & $+8.932087551\cdot 10^{-7}$ & $+3.205282889\cdot 10^{-6}+5.612196126\cdot 10^{-6}\,i$&\\
 \hline
 $A_{20}$ & $-4.235989659\cdot 10^{-8}$ & $-1.659620988\cdot 10^{-7}-2.661550798\cdot 10^{-7}\,i$ & $+8.1249(2)\cdot 10^{-7}-8.72727(4)\cdot 10^{-7}\,i$\\
          Pred. & $-4.235989677\cdot 10^{-8}$ & $-1.659621000\cdot 10^{-7}-2.661550810\cdot 10^{-7}\,i$&\\
 \hline
\end{tabular}
}
\end{center}
\newpage
In the following table, we compare the $1/\epsilon^2$ and $1/\epsilon$ poles of the 2-loop form factors against predicted IR poles (\ref{eq:isoft}, \ref{eq:icollinear}) as well as provide the $\epsilon^0$ term (before IR subtraction) for a point in the physical region with $s/m_t^2=142/17$, $t/m_t^2=-125/22$, $m_Z^2/m_t^2=5/18$, $m_t=1$. We only note here that we observe an improved agreement for the physical combinations of these form factors. The digits in parentheses for the $\epsilon^0$ term denote the uncertainty in the last digit.
\begin{center}
{\tiny
\begin{tabular}{ |c|c|c|c| } 
 \hline
 FF & $1/\epsilon^2$ & $1/\epsilon$ & $\epsilon^0$\\
 \hline\hline
 $A_1$ & $-5.726898 \cdot 10^{-1}-4.634791 \cdot 10^{-1}i$ & $-6.75706 \cdot 10^{-1}-4.05460\,i$ & $6.87787(1) -7.90340(1)\,i$\\
       Pred.& $-5.726897\cdot10^{-1}-4.634791\cdot10^{-1}i$ & $-6.75704 \cdot 10^{-1}-4.05460\,i$ &\\
 \hline
 $A_2$ & $+4.153857\cdot10^{-1}+1.097935\cdot10^{-1}i$ & $+1.40864+2.02204\,i$ & $-2.53566(2) +7.06651(3) \,i$\\
       Pred.& $+4.153857\cdot10^{-1}+1.097934\cdot10^{-1}i$ & $+1.40865+2.02204\,i$ &\\
 \hline
 $A_3$ & $+2.003102\cdot10^{-1}+3.116062\cdot10^{-1}i$ & $-5.02052\cdot10^{-1}+1.86425\,i$ & $-3.99592(2) +2.59711(2) \,i$\\
       Pred.& $+2.003101\cdot10^{-1}+3.116062\cdot10^{-1}i$ & $-5.02053\cdot10^{-1}+1.86425\,i$ &\\
 \hline
 $A_4$ & $+3.147592\cdot10^{-2}+9.237206\cdot10^{-4}i$ & $+1.39272\cdot10^{-1}+1.16086\cdot10^{-1}i$ & $-4.1039(4)\cdot 10^{-2}+5.40365(5)\cdot 10^{-1}\,i$\\
       Pred.& $+3.147591\cdot10^{-2}+9.237121\cdot10^{-4}i$ & $+1.39272\cdot10^{-1}+1.16086\cdot10^{-1}i$ &\\
 \hline
 $A_5$ & $+1.041667\cdot10^{-1}+5.382124\cdot10^{-2}i$ & $+2.44023\cdot10^{-1}+5.97453\cdot10^{-1}i$ & $-8.96421(5)\cdot 10^{-1}+ 1.736695(6) \,i$\\
       Pred.& $+1.041667\cdot10^{-1}+5.382123\cdot10^{-2}i$ & $+2.44022\cdot10^{-1}+5.97453\cdot10^{-1}i$ &\\
 \hline
 $A_6$ & $+1.242527\cdot10^{-1}+6.941130\cdot10^{-2}i$ & $+2.52191\cdot10^{-1}+7.24307\cdot10^{-1}i$ & $-1.20930(2)+1.93865(2) \,i$\\
       Pred.& $+1.242527\cdot10^{-1}+6.941131\cdot10^{-2}i$ & $+2.52189\cdot10^{-1}+7.24307\cdot10^{-1}i$ &\\
 \hline
 $A_7$ & $+3.147592\cdot10^{-2}+9.237206\cdot10^{-4}i$ & $+1.39272\cdot10^{-1}+1.16086\cdot10^{-1}i$ & $-4.1039(4)\cdot 10^{-2}+5.40365(4)\cdot 10^{-1}\,i$\\
       Pred.& $+3.147591\cdot10^{-2}+9.237121\cdot10^{-4}i$ & $+1.39272\cdot10^{-1}+1.16086\cdot10^{-1}i$ &\\
 \hline
 $A_8$ & $-1.017708\cdot10^{-2}+8.808524\cdot10^{-2}i$ & $-4.41618\cdot10^{-1}+2.61228\cdot10^{-1}i$ & $-1.00384(5) -4.4284(4)\cdot 10^{-1}\,i$\\
       Pred.& $-1.017707\cdot10^{-2}+8.808519\cdot10^{-2}i$ & $-4.41613\cdot10^{-1}+2.61225\cdot10^{-1}i$ &\\
 \hline
 $A_9$ & $+7.168287\cdot10^{-2}-5.063902\cdot10^{-2}i$ & $+5.37076\cdot10^{-1}+9.24698\cdot10^{-2}i$ & $3.07426(8)\cdot 10^{-1}1.266108(9) \,i$\\
       Pred.& $+7.168286\cdot10^{-2}-5.063902\cdot10^{-2}i$ & $+5.37075\cdot10^{-1}+9.24707\cdot10^{-2}i$ &\\
 \hline
 $A_{10}$ & $+1.873343\cdot10^{-2}-8.497011\cdot10^{-2}i$ & $+4.70733\cdot10^{-1}-2.17284\cdot10^{-1}i$ & $+9.3643(1) \cdot 10^{-1} + 6.3029(1)\cdot 10^{-1}\,i$\\
          Pred.& $+1.873344\cdot10^{-2}-8.497010\cdot10^{-2}i$ & $+4.70734\cdot10^{-1}-2.17286\cdot10^{-1}i$ &\\
 \hline
 $A_{11}$ & $-7.675742\cdot10^{-2}+5.097567\cdot10^{-2}i$ & $-5.57824\cdot10^{-1}-1.06514\cdot10^{-1}i$ & $-3.1397(3) \cdot 10^{-1} -1.35727(4) \,i$\\
          Pred.& $-7.675741\cdot10^{-2}+5.097571\cdot10^{-2}i$ & $-5.57827\cdot10^{-1}-1.06513\cdot10^{-1}i$ &\\
 \hline
 $A_{12}$ & $+7.675742\cdot10^{-2}-5.097567\cdot10^{-2}i$ & $+5.57824\cdot10^{-1}+1.06514\cdot10^{-1}i$ & $+3.1397(3) \cdot 10^{-1} +1.35727(4)\,i$\\
          Pred.& $+7.675741\cdot10^{-2}-5.097571\cdot10^{-2}i$ & $+5.57827\cdot10^{-1}+1.06513\cdot10^{-1}i$ &\\
 \hline
 $A_{13}$ & $-1.873343\cdot10^{-2}+8.497011\cdot10^{-2}i$ & $-4.70733\cdot10^{-1}+2.17284\cdot10^{-1}i$ & $-9.3644(1) \cdot 10^{-1} - 6.3029(1) \cdot 10^{-1}\,i$\\
          Pred.& $-1.873344\cdot10^{-2}+8.497010\cdot10^{-2}i$ & $-4.70734\cdot10^{-1}+2.17286\cdot10^{-1}i$ &\\
 \hline
 $A_{14}$ & $-7.168287\cdot10^{-2}+5.063902\cdot10^{-2}i$ & $-5.37076\cdot10^{-1}-9.24698\cdot10^{-2}i$ & $-3.07426(8)\cdot 10^{-1} -1.266108(9) \,i$\\
          Pred.& $-7.168286\cdot10^{-2}+5.063902\cdot10^{-2}i$ & $-5.37075\cdot10^{-1}-9.24707\cdot10^{-2}i$ &\\
 \hline
 $A_{15}$ & $+1.017708\cdot10^{-2}-8.808524\cdot10^{-2}i$ & $+4.41618\cdot10^{-1}-2.61228\cdot10^{-1}i$ & $1.00384(4) + 4.4283(4)\cdot 10^{-1}\,i$\\
          Pred.& $+1.017707\cdot10^{-2}-8.808519\cdot10^{-2}i$ & $+4.41613\cdot10^{-1}-2.61225\cdot10^{-1}i$ &\\
 \hline
 $A_{16}$ & $-6.195421\cdot10^{-2}-9.197693\cdot10^{-2}i$ & $+1.25592\cdot10^{-1}-6.06299\cdot10^{-1}i$ & $1.76383(3) -9.4291(3) \cdot 10^{-1} \,i$\\
          Pred.& $-6.195417\cdot10^{-2}-9.197695\cdot10^{-2}i$ & $+1.25596\cdot10^{-1}-6.06299\cdot10^{-1}i$ &\\
 \hline
 $A_{17}$ & $+9.152404\cdot10^{-4}+4.922399\cdot10^{-3}i$ & $-1.47185\cdot10^{-2}+2.71477\cdot10^{-2}i$ & $-8.6390(6)\cdot 10^{-2}+2.7504(7)\cdot 10^{-2}\,i$\\
          Pred.& $+9.152368\cdot10^{-4}+4.922402\cdot10^{-3}i$ & $-1.47187\cdot10^{-2}+2.71472\cdot10^{-2}i$ &\\
 \hline
 $A_{18}$ & $+6.800443\cdot10^{-3}+5.687424\cdot10^{-3}i$ & $+7.80438\cdot10^{-3}+4.98318\cdot10^{-2}i$ & $-1.02182(8)\cdot 10^{-1}+1.37512(8)\cdot 10^{-1}\,i$\\
          Pred.& $+6.800439\cdot10^{-3}+5.687435\cdot10^{-3}i$ & $+7.80405\cdot10^{-3}+4.98315\cdot10^{-2}i$ &\\
 \hline
 $A_{19}$ & $+4.208648\cdot10^{-3}+4.547692\cdot10^{-3}i$ & $-3.01730\cdot10^{-4}+3.55035\cdot10^{-2}i$ & $-7.895(10)\cdot 10^{-2}+7.980(11)\cdot 10^{-2}\,i$\\
          Pred.& $+4.208616\cdot10^{-3}+4.547808\cdot10^{-3}i$ & $-3.13880\cdot10^{-4}+3.55067\cdot10^{-2}i$ &\\
 \hline
 $A_{20}$ & $+9.152403\cdot10^{-4}+4.922399\cdot10^{-3}i$ & $-1.47185\cdot10^{-2}+2.71477\cdot10^{-2}i$ &  $-8.6391(6)\cdot 10^{-2}+2.7504(7)\cdot 10^{-2}\,i$\\
          Pred.& $+9.152368\cdot10^{-4}+4.922402\cdot10^{-3}i$ & $-1.47187\cdot10^{-2}+2.71472\cdot10^{-2}i$ &\\
 \hline
\end{tabular}
}
\end{center}

\newpage
\section{Subtraction scheme dependence}
\label{sec:altsubtraction}
In the main text of this article we define the finite remainders for the amplitudes according to the ``$q_\mathrm{T}$ scheme'' \cite{Catani:2013tia}; see section~\ref{sec:scheme}.
As described in section~\ref{sec:results}, the choice of scheme can significantly affect the results and the level of agreement with the available approximations.
In this appendix we demonstrate this effect explicitly by presenting a selection of our results using an alternative definition of the finite remainders in Catani's original scheme \cite{Catani:1998bh}.

At the level of form factors, the finite remainders in Catani's original scheme are obtained from their ``$q_\mathrm{T}$ scheme'' analog in \eqref{eq:finiteff} according to
\begin{align}
   A_i^{(1),\text{fin},\text{Catani}}  &= A_i^{(1),\text{fin}},\\
\label{eq:ffcatani}
   A_i^{(2),\text{fin},\text{Catani}}  &= A_i^{(2),\text{fin}} + \Delta I_1 A_i^{(1),\text{fin}},\\
   \intertext{where}
\label{eq:transformtocatani}
   \Delta I_1 &= -\frac{1}{2}\pi^2 C_A +i \pi \beta_0,
\end{align}
see also eqs.\ 4.9 and 4.10 in \cite{vonManteuffel:2015msa}.
The transformation of the helicity amplitudes follows the same pattern. For the interference terms considered in eqs.\ \eqref{eq:interf1loop} and \eqref{eq:interf2loop} an additional factor of 2 needs to be taken into account  and the term due to $i\pi\beta_0$ does not contribute.

\begin{figure}[th]
\centering
\includegraphics[height=48mm]{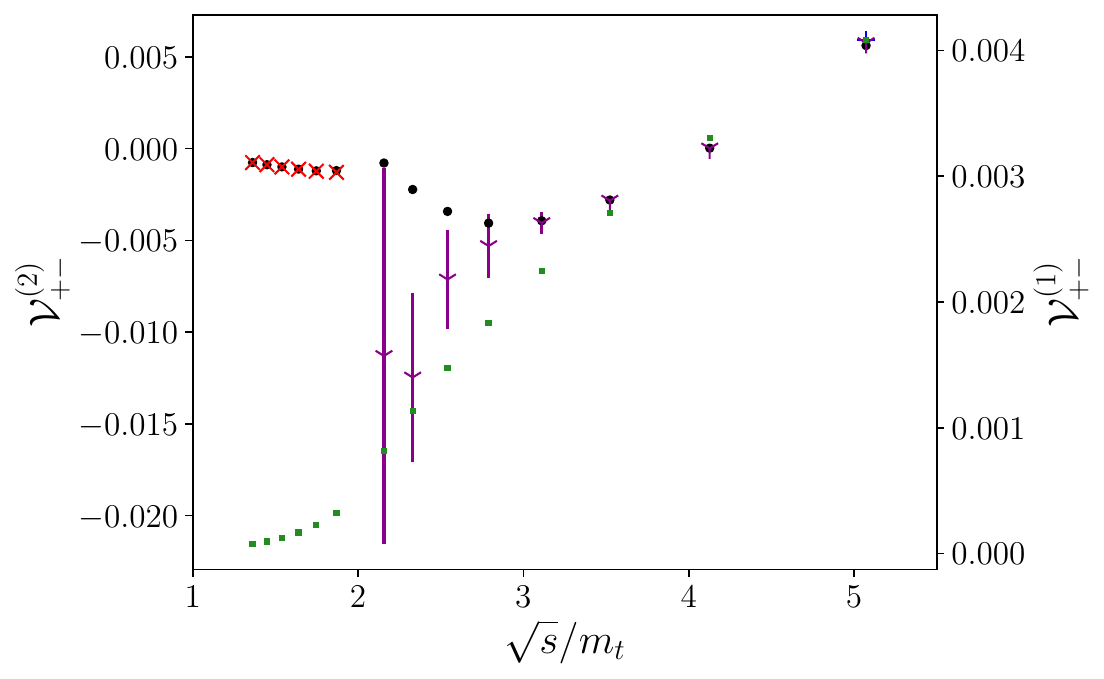}
\includegraphics[height=48mm]{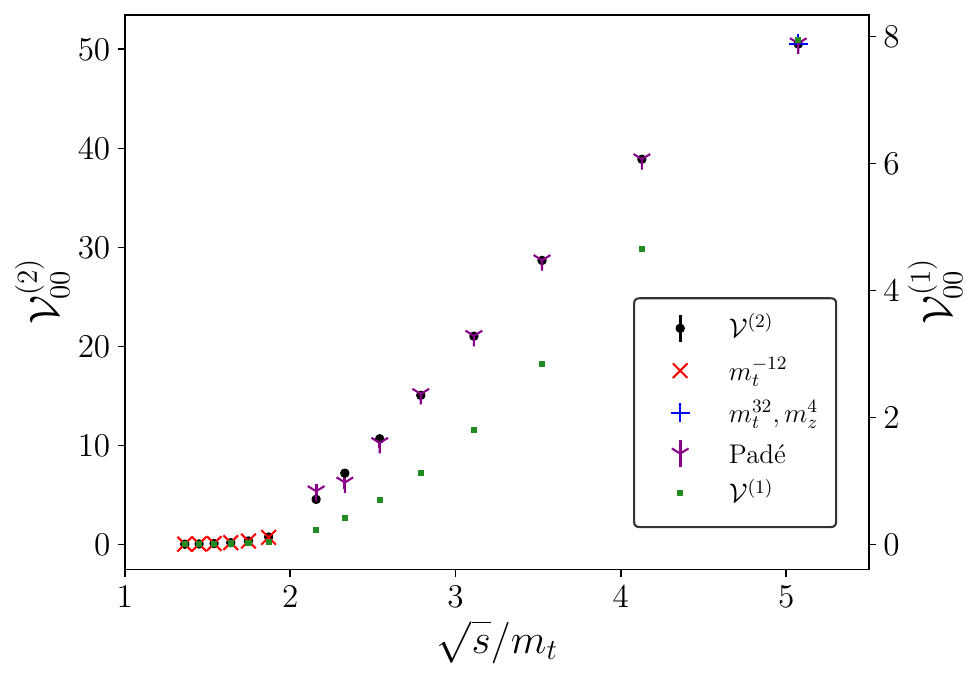}
\caption{The $\sqrt{s}$ dependence of 1-loop and 2-loop interferences for polarised $ZZ$ production in gluon fusion at $\cos(\theta)=-0.1286$. 
Here we reproduce the top left and bottom right panels of figure~\ref{fig:helampsbeta} using Catani's original subtraction scheme \cite{Catani:1998bh}.}
\label{fig:catanihelampsbeta}
\end{figure}

\begin{figure}[th]
\centering
\includegraphics[height=47mm]{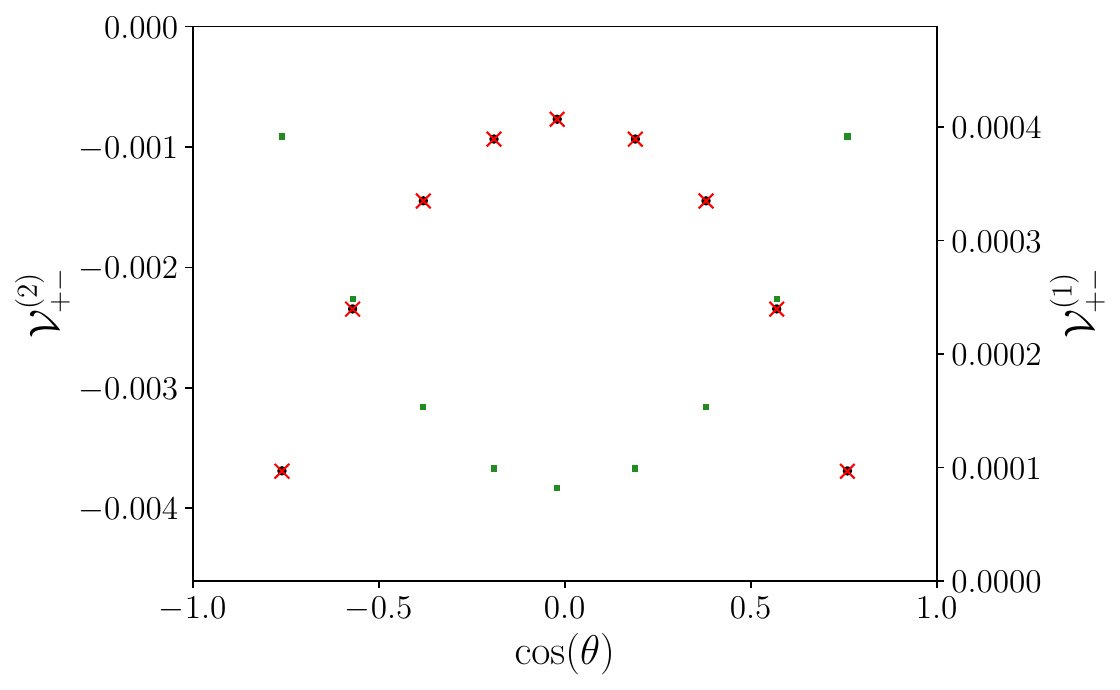}
\includegraphics[height=47mm]{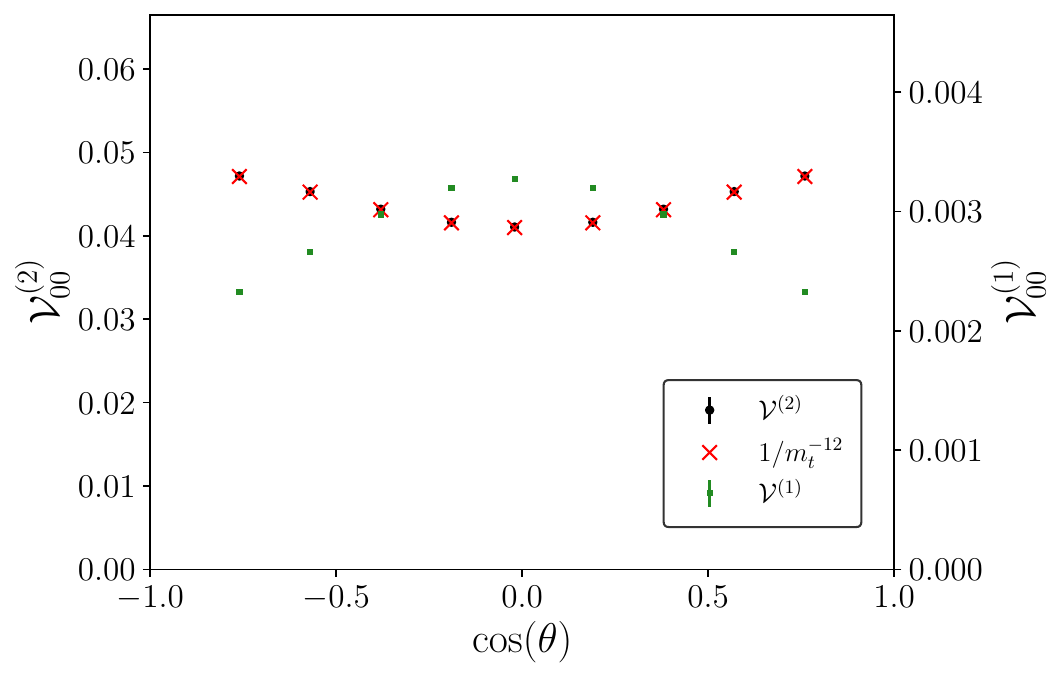}
\caption{The $\cos(\theta)$ dependence of 1-loop and 2-loop interferences for polarised $ZZ$ production in gluon fusion at $\sqrt{s}/m_t=1.426$. The large top-quark mass expansion~\cite{Davies:2020lpf} (to order $1/m_t^{12}$) is shown for comparison. 
Here we reproduce the top left and bottom right panels of figure~\ref{fig:helampsthetahtl} using Catani's original subtraction scheme \cite{Catani:1998bh}.}
\label{fig:catanihelampsthetahtl}
\end{figure}

\begin{figure}[th]
\centering
\includegraphics[height=48mm]{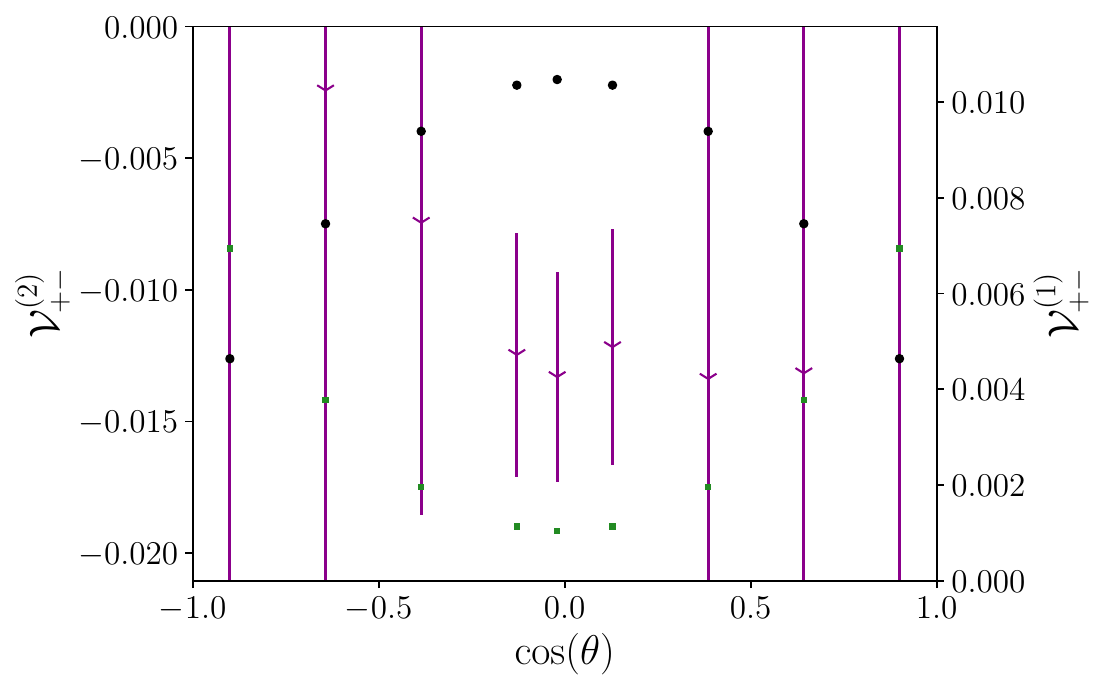}
\includegraphics[height=48mm]{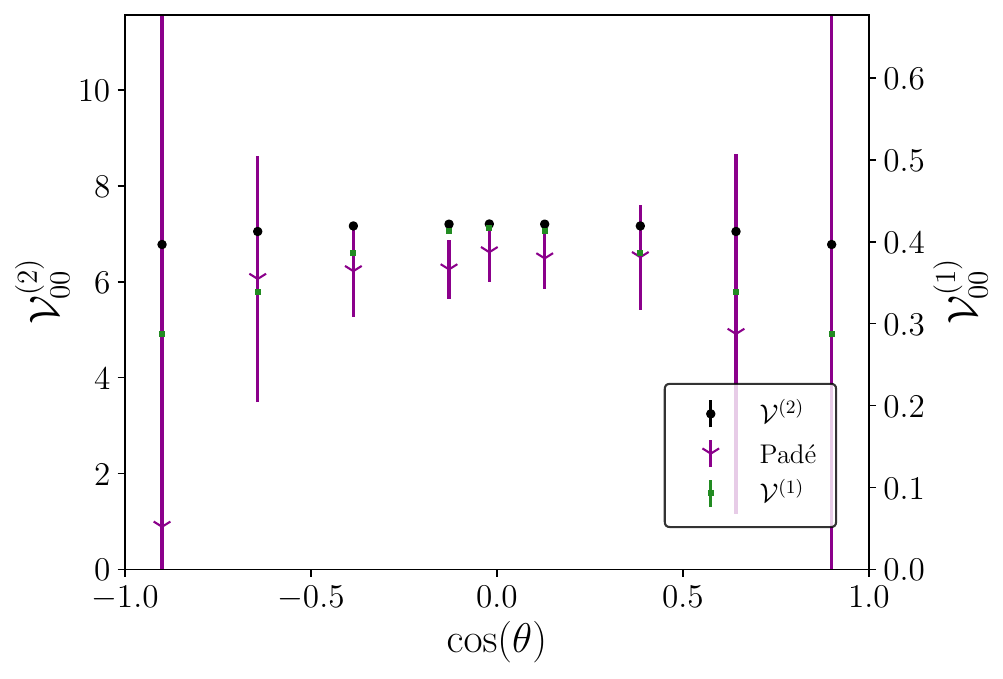}
\caption{The $\cos(\theta)$ dependence of 1-loop and 2-loop interferences for polarised $ZZ$ production in gluon fusion at $\sqrt{s}/m_t=2.331$. The Pad\'{e} improved small top-quark mass expansion~\cite{Davies:2020lpf} is shown for comparison. 
Here we reproduce the top left and bottom right panels of figure~\ref{fig:helampsthetamid} using Catani's original subtraction scheme \cite{Catani:1998bh}.}
\label{fig:catanihelampsthetamid}
\end{figure}

\begin{figure}[th]
\centering
\includegraphics[height=48mm]{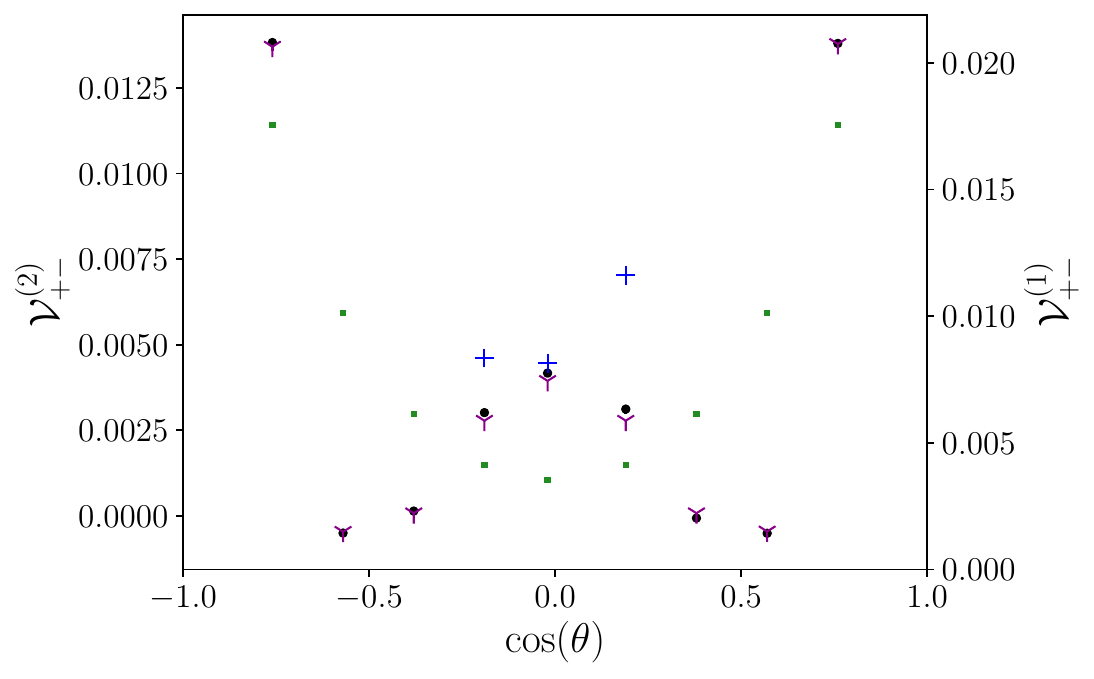}
\includegraphics[height=48mm]{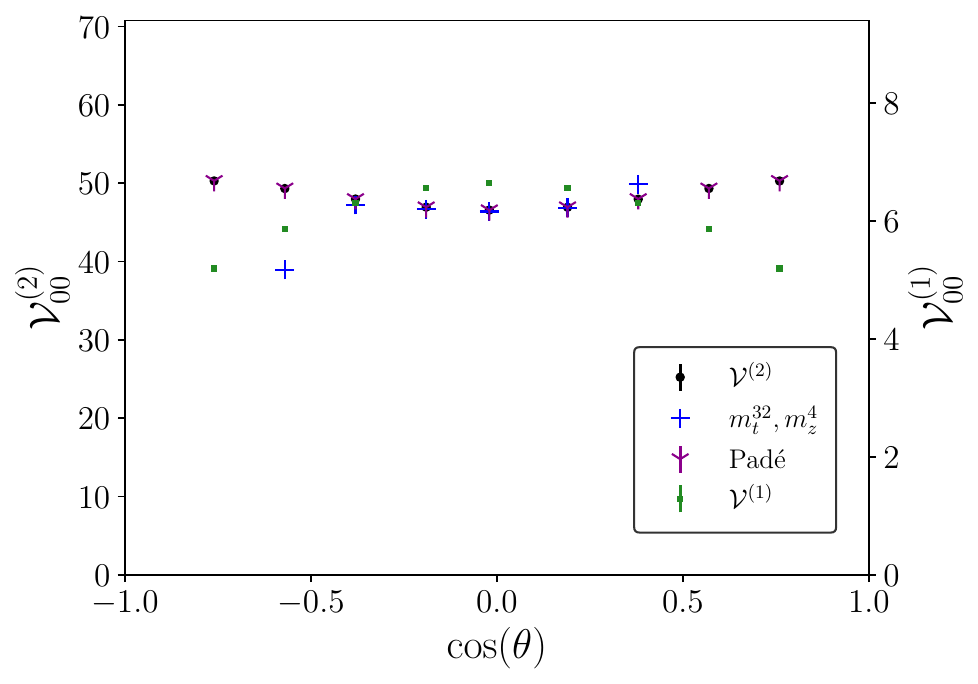}
\caption{The $\cos(\theta)$ dependence of 1-loop and 2-loop interferences for polarised $ZZ$ production in gluon fusion at $\sqrt{s}/m_t=4.703$. The small top-quark mass expansion (to order $m_t^{32}$) and Pad\'{e} improved expansion~\cite{Davies:2020lpf} are shown for comparison. 
Here we reproduce the top left and bottom right panels of figure~\ref{fig:helampsthetahe} using Catani's original subtraction scheme \cite{Catani:1998bh}.}
\label{fig:catanihelampsthetahe}
\end{figure}

As can be seen in figures \ref{fig:catanihelampsbeta}, \ref{fig:catanihelampsthetahtl}, \ref{fig:catanihelampsthetamid}, and \ref{fig:catanihelampsthetahe}, the 2-loop corrections can show a rather different qualitative behaviour than the 1-loop corrections in Catani's original scheme.
This is in contrast to the corresponding results in the ``$q_\mathrm{T}$ scheme'' in figures \ref{fig:helampsbeta}, \ref{fig:helampsthetahtl}, \ref{fig:helampsthetamid}, and \ref{fig:helampsthetahe}.
Moreover, the relative agreement between the expansion results and our exact calculation depends greatly on the choice of scheme for the finite remainder; it is significantly better in the ``$q_\mathrm{T}$ scheme'' than in Catani's original scheme.
In order to assess which relative error reflects better the resulting relative error on physical observables, the corresponding real radiation contributions would need to be taken into account in the respective scheme as well.

\newpage
\bibliographystyle{JHEP}
\bibliography{refs.bib}
\end{document}